\documentclass[a4paper,11pt]{article}
\pdfoutput=1
\usepackage{float}
\usepackage{jheppub}
\usepackage{amsmath,amsfonts,amssymb,amsthm,graphicx,color}
\usepackage{mathrsfs}
\usepackage[dvipsnames]{xcolor}
\usepackage{verbatim}
\usepackage{enumerate}
\usepackage{epsfig}
\usepackage{multirow}
\usepackage{comment}
\usepackage{datetime}
\usepackage{slashed}
\usepackage{framed}
\usepackage{longtable}
\usepackage[mathscr]{euscript}
\usepackage{cancel}
\usepackage{tensor}
\usepackage{mathtools}
\usepackage{caption}
\usepackage{subcaption}
\usepackage[multiple]{footmisc}
\usepackage{pgfplots}
\usepackage[T1]{fontenc}
\usepackage[utf8]{inputenc}
\usepackage{tikz}
\usepackage{placeins}
\usepackage{hyperref}
\usetikzlibrary{decorations.markings}
\usepackage{multirow}
\usepackage{graphicx}

\def\ii{{\rm i}}
\newcommand{\ft}[2]{{\textstyle\frac{#1}{#2}}}
\newcommand{\nn}{\nonumber}
\newcommand{\diff}{\mathrm{d}}
\newcommand{\odd}{\mathrm{odd}}
\newcommand{\ev}{\mathrm{ev}}
\newcommand{\inn}{\mathrm{in}}
\newcommand{\out}{\mathrm{out}}
\newcommand{\E}{\mathcal{E}}
\newcommand{\T}{\mathcal{T}}

\def\be{\begin{equation}}
	\def\ee{\end{equation}}
\def\bea{\begin{align}}
	\def\eea{\end{align}}
\def\beaq{\begin{eqnarray}}
	\def\eeaq{\end{eqnarray}}

\title{Gravitational waveforms from binaries in higher-derivative gravity: a Love story}
\author[a]{Pablo A.~Cano,}
\author[b]{Francesco Fucito,}
\author[b]{Jose F.~Morales,}
\author[c]{Alejandro Ruip\'erez}
\affiliation[a]{Departamento de F\'isica, Universidad de Murcia, Campus de Espinardo, 30100 Murcia, Spain}
\vspace{5mm}
\affiliation[b]{Dipartimento di Fisica, Universit\`a di Roma ``Tor Vergata" \& Sezione INFN  ``Roma Tor Vergata", Via della Ricerca Scientifica 1, 00133, Roma, Italy}
\affiliation[c]{Dipartimento di Fisica e Astronomia ``Galileo Galilei'', Universit\`a di Padova \& INFN Sezione di Padova, Via Marzolo 8, 35135, Padova, Italy}

\abstract{
We study the emission of gravitational waves by a test particle orbiting a non-rotating black hole in higher-derivative gravity theories with cubic and quartic contractions of the Riemann tensor. To this aim, we first derive the master equations describing even- and odd-parity perturbations in the presence of an arbitrary source term,
and then construct a Post-Minkowskian expansion of the solutions to the homogeneous master equations. Specializing to a circular binary system, we compute the Post-Newtonian expansion of the waveform, as well as the energy and angular-momentum fluxes at infinity. We show that higher-derivative corrections to the waveform and to the fluxes always appear at 5PN order, and are universally proportional to the Love number describing the deformability of the geometry under the $\ell=2$ mode perturbation. These analytical results are validated against numerical computations, which also allow us to extend the analysis to larger velocities.
}

\begin{document}
	\maketitle
	\flushbottom

    \newpage
	
\section{Introduction and summary of results}

The  experimental detection of gravitational waves (GW) by the LIGO/VIRGO collaboration \cite{LIGOScientific:2016aoc,TheLIGOScientific:2016src,LIGOScientific:2018mvr,LIGOScientific:2019fpa,Abbott:2020tfl,Abbott:2020mjq,Abbott:2020khf,LIGOScientific:2020tif} has opened a new era of GW astronomy.
Next generation detectors \cite{LISA:2017pwj,LISA:2022kgy,ET:2019dnz,Punturo:2010zz} will significantly boost the frequency range and improve the sensitivity of the experimental set-ups providing an unprecedented opportunity to test general relativity (GR) in the strong field regime.

The gravitational-wave signal produced by the coalescence of a binary system provides one of the cleanest signatures of the nature of the colliding objects and of their interactions. This information is encoded in the inspiral–merger–ringdown waveform. In the case of black hole coalescence, the waveform is completely determined by a handful of parameters according to the vacuum Einstein's field equations. Any deviation from the expected signal would indicate either a departure from the compactness of black holes or a modification of GR itself.

The nature of compact objects can reveal itself during the inspiral phase both through tidal deformations \cite{Cardoso:2017cfl,Chakraborty:2026qru} and through deviations from the expected multipolar structure \cite{Bianchi:2020miz}, and via gravitational-wave echoes during the ringdown \cite{Cardoso:2016oxy,ToVSapQNM}. On the other hand, deviations from GR  typically modify the properties of the signal quantitatively. For instance, they affect the Post-Newtonian (PN) coefficients that characterize the inspiral waveform in the small-velocity expansion \cite{Blanchet:2013haa} and introduce shifts in the black hole quasinormal mode frequencies \cite{Berti:2025hly,Antoniou:2024jku}. However, beyond-GR theories also introduce qualitative differences with respect to GR, such as non-zero tidal deformability for black holes \cite{Cardoso:2018ptl,Cai:2019npx,DeLuca:2022tkm,Katagiri:2024fpn,Barbosa:2025uau,Cano:2025zyk,Wang:2026qst} and the breaking of quasinormal mode isospectrality \cite{Cano:2021myl,Li:2023ulk,Cano:2024wzo}.

In this paper we explore  modifications of GR (a somewhat different approach was followed in \cite{Bianchi:2023sfs, Cipriani:2024ygw,Bianchi:2024vmi}). A natural and systematic framework to parametrize deviations with respect to GR is provided by effective field theory (EFT). EFT has long been used in theoretical physics as an approach to capture the low-energy effects of a hypothetical UV theory in an agnostic way \cite{Polchinski:1983gv,Lepage:1989hf}. In the case of gravity, the size of the higher-derivative corrections is set by the ratio between the curvature scale and a certain scale of new physics \cite{Baker:2014zba}. At the level of the Solar System, the maximum curvature scales are of order $10^{-8}\,\mathrm{km}^{-1}$. This corresponds to the weak-field regime, where so far experiments have shown no deviations from GR \cite{Will:2005va}. By contrast, current GW observations probe curvature scales of order $1\mathrm{km}^{-1}$, corresponding to the strong-field regime of highly compact objects, which remains a largely uncharted territory.

Without introducing additional light degrees of freedom other than the metric, and up to order eight in derivatives, the most general EFT extension of GR up to field redefinitions was presented in \cite{Endlich:2017tqa}.\footnote{Causality constraints on higher-derivative gravity theories have been extensively studied in the past literature, starting with \cite{Gruzinov:2006ie,Camanho:2014apa}.} The corresponding effective action contains six- and eight-derivative terms constructed out of certain cubic and quartic contractions of the Riemann tensor. Further demanding invariance under parity, the explicit form of the action reads
\begin{equation}\label{eq:EFTofGR}
S_{\rm EFT}\,=\, \frac{1}{2\kappa^2}\int \diff^4x\sqrt{-g}\left[R+\lambda_{3} \, {\cal C}_{3}+\lambda_4\, \mathcal{C}_{2}^2+{\tilde\lambda}_4\, \tilde{\mathcal{C}}_{2}^2 +\dots \right]\, ,
\end{equation}
where $\lambda_3, \lambda_4$ and ${\tilde\lambda}_4$ are dimensionful higher-derivative coupling constants and
\begin{equation}\label{eq:defCs}
{\cal C}_{3}\,=\,\tensor{R}{_{\mu\nu}^{\rho\sigma}}\tensor{R}{_{\rho\sigma }^{\delta\gamma}}\tensor{R}{_{\delta\gamma }^{\mu\nu }}\,, \hspace{1cm}\mathcal{C}_{2}=R_{\mu\nu\rho\sigma} R^{\mu\nu\rho\sigma}\, , \hspace{1cm} \tilde{\mathcal{C}}_{2}=R_{\mu\nu\rho\sigma} \tilde{R}^{\mu\nu\rho\sigma}\, ,
\end{equation}
being ${\tilde R}^{\mu\nu\rho\sigma}=\frac{1}{2}\epsilon^{\mu\nu\alpha\beta}\tensor{R}{_{\alpha\beta}^{\rho\sigma}}$ the dual of the Riemann tensor. The dots in \eqref{eq:EFTofGR} denote terms of higher order in derivatives, which are assumed to be subleading with respect to the ones we include.\footnote{The corrections coming from a term with $2n$ derivatives will be effectively controlled by a dimensionless parameter $\lambda_{n} {\cal R}^n$, where $\cal R$ is a characteristic curvature scale and $\lambda_{n}$ is a higher-derivative coupling constant with dimensions of length${}^{2n-2}$. If we assume that there is only one relevant scale of new physics controlling the whole higher-derivative expansion, and denote it by $\ell_{\rm new}$, the relative correction to GR would be of the order of $\lambda_{n}{\cal R}^{n-1}\sim \left(\ell^2_{\rm new}{\cal R}\right)^{n-1}$, and therefore terms with a higher number of derivatives will be more and more suppressed.}

Black hole solutions of the effective action \eqref{eq:EFTofGR} were constructed in \cite{Cardoso:2018ptl,Cano:2019ore}, working at linear order in the higher-derivative couplings. The aim of this paper is to investigate how these higher-derivative interactions modify gravitational-wave signals. More concretely, we will consider an extreme-mass-ratio inspiral binary and compute the corrections to the Post-Newtonian (PN) expansion of the waveform. To this end, we will model the object with lower mass as a perturbation around the primary black hole and employ the framework of black hole perturbation theory. Partial results obtained using different techniques are already available in the literature; the effects of quartic interactions were studied in \cite{Endlich:2017tqa,Cardoso:2018ptl,Sennett:2019bpc}, the cubic terms have been recently explored in \cite{Brandhuber:2024bnz,Brandhuber:2024lgl,Liu:2024atc} and quadratic corrections were studied in \cite{Antoniou:2024jku,Antoniou:2026nhh}. Here, we investigate both cubic and quartic interactions.

In Einstein gravity, first-order perturbations around Schwarzschild split into even- and odd-parity sectors, described respectively by the Regge–Wheeler and Zerilli master equations \cite{Regge:1957td,Zerilli:1970se}, see e.g.~\cite{Sago:2002fe,Nakano:2003he} for a review. These results were later reformulated by Teukolsky in a more compact way using the Newman–Penrose formalism \cite{Teukolsky:1972my}. A PN expansion of the solutions to the Teukolsky equation was developed in \cite{Poisson:1993vp,Cutler:1993vq,Tagoshi:1993dm,Sasaki:1994rw,Tagoshi:1994sm,Shibata:1994jx,Poisson:1995vs,Mano:1996vt}, in what is now known as the MST method, see \cite{Mino:1997bx} for a review. These results have recently been re-derived in \cite{Fucito:2023afe,Cipriani:2025ikx} using the quantum Seiberg–Witten/gravity correspondence \cite{Aminov:2020yma,Bianchi:2021xpr,Bianchi:2021mft,Bonelli:2021uvf,Bonelli:2022ten,Consoli:2022eey,Bautista:2023sdf,Aminov:2023jve,ToVSapQNM,Fucito:2023afe,Cipriani:2025ikx}. These techniques are flexible and also apply to cosmology
\cite{Bianchi:2024mlq} and gravitational scattering processes \cite{Fucito:2024wlg,Cipriani:2026myb}. Here we extend them to higher-derivative gravity.

Let us summarize our results. First, we derive the corrected master equations governing the dynamics of odd and even perturbations around a static black hole in the presence of an arbitrary source. Next, we specialize to a point-like source of mass $\mu$, moving on a circular orbit of radius $r_0\,=\,M/v^2$ around a black hole of mass $M\gg \mu$, and compute the corresponding waveform, as well as the energy and angular momentum fluxes at infinity. At the technical level, this is done by adapting the approach of \cite{Cipriani:2025ikx} to construct the solutions of the master equations in a Post-Minkowskian (PM) expansion, which describes the gravitational wave in the \emph{near-zone} $\omega M\ll \omega r \ll 1$ where the inspiral takes place.  We present our results up to 6PN order, which allows us to capture the leading and next-to-leading contributions to the waveform and fluxes from the higher-derivative terms, as we discuss in the following paragraph.

The corrections to the waveform can be classified into two different sets. On the one hand, there is a first set that arises after imposing appropriate boundary conditions (incoming) at the horizon. As it turns out, these appear always at $(2\ell+1)$PN\footnote{$\ell$ is the orbital angular momentum number.} order and are universally proportional to the static Love numbers of the corrected black hole geometry. Then, there is a second set of contributions coming from the corrections to the geodesic motion and from the explicit corrections to the master equations (in particular, to the effective potential). The order at which the latter appear depends on the specific form of the higher-derivative terms. For instance, the cubic corrections start contributing at 6PN, while the quartic ones at 8PN  (so we will neglect them when presenting our final results). Consequently, we have that the leading corrections to the waveform for the lowest $\ell=2$ mode come from the first set, and appear at 5PN for both cubic and quartic theories. This can be summarized in a rather compact form by writing our final formula for the (modulus of the) waveform, $|\tilde Z_{\ell m}|$. To this aim, we split the GR and higher-derivative contributions as follow
\begin{equation}
|\tilde Z_{\ell m}|\,=\,|\tilde Z_{\ell m}|_N \left(\Gamma_{{\rm GR}, \ell m}+\delta \Gamma_{\ell m}\right)\,.
\end{equation}
Here $|\tilde Z_{\ell m}|_N$ and $\Gamma_{{\rm GR}, \ell m}$ stand for leading contribution and the PN corrections in GR, which are given in the main text. Higher-derivative interactions modify the PN expansion of the waveform, and their contributions are encoded in the quantity $\delta \Gamma_{\ell m}$. Up to 6PN order, our result for the even and odd sectors is the following:
\begin{equation}
\begin{aligned}
\delta\Gamma^{\ev}_{\ell m}\,=\,& 2k_2^{+} v^{10}\left(1+\frac{3}{2}v^2+\dots\right)\delta_{\ell 2}\delta_{m2}+\frac{2 \ell (43+8 \ell) v^{12}}{-5+2 \ell}\frac{\lambda_3}{M^4}+\dots\,, \\[1mm]
\delta\Gamma^{\odd}_{\ell m}\,=\,&-3k_2^-  \,v^{10}\left(1+\frac{8v^2}{3}+\dots\right)\delta_{\ell 2}\delta_{m1}-\frac{10 (1+\ell) (-29+8 \ell) v^{12}}{-5+2 \ell}\frac{\lambda_3}{M^4}+\dots\,,
\end{aligned}
\end{equation}
where
\begin{equation}
k_2^+\,=\,28\frac{\lambda_3}{M^4}+\frac{1008}{25}\frac{\lambda_4}{M^6}\,,\hspace{1cm}
k_2^- \,=\,-20\frac{\lambda_3}{M^4}-\frac{432}{25}\frac{\lambda_4}{M^6}+\frac{96}{5}\frac{{\tilde\lambda}_4}{M^6}
\end{equation}
are the static tidal Love numbers \cite{Cardoso:2018ptl,Cai:2019npx,DeLuca:2022tkm,Katagiri:2024fpn,Barbosa:2025uau,Cano:2025zyk} in the conventions of \cite{Cano:2025zyk}. A main implication is that the leading contribution from the higher-derivative interactions to the energy and angular momentum fluxes at infinity is proportional to the Love number $k_2^+$ associated to the dominant $\ell=2$ parity-even mode. This is reflected in our final formula for the energy flux:
\begin{equation}
\frac{\diff {\cal E}}{\diff t}\,=\,\frac{32\mu^2v^{10}}{5M^2}\left(\eta^{\rm GR}+\delta \eta\right)\,,
\end{equation}
where
\begin{equation}\label{deltaetaintro}
\begin{aligned}
\eta^{\rm GR}\,=\,& 1-\frac{1247 v^2}{336}+4
   \pi  v^3-\frac{44711 v^4}{9072}+\ldots \,,\\[1mm]
\delta\eta\,=\,& 4 k_2^+ v^{10}-v^{12} \left(\frac{k_2^-}{6}+ \frac{88k_2^+}{21} +\frac{612\lambda_3}{M^4}\right)+\dots \,.
\end{aligned}
\end{equation}

The plan of the paper is as follows:
in section~\ref{sec:BHsoln} we review few important aspects of the effective field theories of gravity we consider in the paper and  their spherically-symmetric black hole solutions. In section~\ref{sec:grav_pert} we find the master equations for the odd and even perturbations, accounting for an arbitrary source term.
In section~\ref{sec:scalar}, mostly for illustrative purposes, we consider scalar perturbations and compute the scalar waveform associated to a charged point-like source on circular motion, as well as the energy flux at infinity. The gravitational counterpart is addressed in section~\ref{sec:waveform+flux} via analytic methods. Then in section~\ref{sec:numericalresults} we test our analytical findings against a numerical approach. Finally, section~\ref{sec:conclusions} contains our conclusions. Lengthy formulas are provided in the appendices and in the ancillary Mathematica file included in the arXiv submission.

\section{Black holes in effective field theories of gravity}\label{sec:BHsoln}

If we do not introduce additional light degrees of freedom, the most general parity-preserving effective field theory (EFT) of gravity in four dimensions up to order eight in derivatives was given in \eqref{eq:EFTofGR}. The equations of motion arising from this effective action can be written as
\begin{equation}\label{eq:fieldeq}
R_{\mu\nu}-\frac{1}{2}g_{\mu\nu} R +{\cal E}_{\mu\nu}\,=\, \kappa^2 T_{\mu\nu}\,,
\end{equation}
where $\kappa^2\,=\, 8\pi G$, and from now on we set $G=1$. $T_{\mu\nu}$ and ${\cal E}_{\mu\nu}$ account for the coupling to matter and for the contribution from the higher-derivative terms, respectively. The  explicit expression of ${\cal E}_{\mu\nu}$ is
\begin{equation}\label{eq:Emunu}
{\cal E}_{\mu\nu}\,=\, P^{\alpha\beta\gamma}{}_{(\mu|}R_{\alpha\beta\gamma|\nu)}-\frac{1}{2}g_{\mu\nu}{\cal L}'-2\nabla^{\alpha}\nabla^{\beta}P_{\alpha(\mu\nu)\beta}  \,,
\end{equation}
where
\begin{equation}
P_{\mu\nu\rho\sigma}\,=\, \frac{\partial {\cal L}'}{\partial R^{\mu\nu\rho\sigma}}\, ,\hspace{1cm} {\cal L}'\,=\,  \lambda_{3}\,\mathcal{C}_3+\lambda_4\, \mathcal{C}_2^2+{\tilde\lambda}_4\, \tilde{\mathcal{C}}_2^2 \,,
\end{equation}
and  ${\cal C}_3, {\cal C}_2$ and $\tilde{\cal C}_2$ were given in \eqref{eq:defCs}.

\subsection{Spherically-symmetric black hole solutions}\label{sec:correctedSchwarzschildsoln}

We are interested in studying perturbations around a static and spherically-symmetric black hole in the EFT \eqref{eq:EFTofGR}. Working linearly in the higher-derivative coupling constants, the solution is given by \cite{Cano:2019ore, Cardoso:2018ptl}
\begin{equation}
\label{eq:bkgmetric}
\diff s^2\,=\, -\,N^2 f\diff t^2 + \frac{\diff r^2}{f} + r^2\, \left(\diff\theta^2+\sin^2\theta\diff\phi^2\right)\, ,
\end{equation}
where
\begin{equation}
\begin{aligned}
f\,=\,&1-\frac{2M}{r}+24\lambda_{3} M^2\left(-\frac{49M}{3r^7}+\frac{9}{r^6}\right)+ 1152\,\lambda_4 M^3\left(\frac{4}{r^9}-\frac{67M}{9r^{10}}\right)\,, \\[1mm]
N\,=\,&1-\frac{108 \lambda_{3}M^2}{r^6}-\frac{1792\lambda_4 M^3}{r^9}\,.
\end{aligned}
\end{equation}
The integration constants have been fixed imposing standard normalization at infinity and that the parameter $M$ still corresponds to the mass of the solution. In contrast, the position of the horizon, which we denoted by $r_+$, is shifted as follows
\begin{equation}\label{eq:rpl}
f(r_+)\,=\,0\,, \hspace{1cm} \Rightarrow \hspace{1cm} r_+\,=\,2M-\frac{5\lambda_3}{8M^3}-\frac{5\lambda_4}{4M^5}\, .
\end{equation}
Finally, we provide the expression for the inverse Hawking temperature $\beta$, as it will enter in some formulas later,
\begin{equation}\label{eq:inversetemperature}
\frac{\beta}{2\pi}\,=\,\frac{1}{2\pi T}\,=\, \frac{2}{N(r_+)f'(r_+)}\,=\,4M \left(1-\frac{\lambda_3}{8 M^4}-\frac{\lambda_4}{2 M^6}\right)\, .
\end{equation}

\subsection{Circular orbits}\label{sec:circulargeo}

 Given that our objective is to investigate the emission of scalar and gravitational waves produced by a point-like source moving along a circular geodesic around the black hole geometry \eqref{eq:bkgmetric}, we devote this section to review how geodesic motion is modified in our EFT, referring to \cite{Cano:2019ore} for further details.

Geodesic motion is governed by the Hamiltonian
\begin{equation}\label{hzero}
 {\cal H}\,=\,\ft12 g^{\mu\nu} P_\mu  P_\nu  =-\frac{\mu^2}{2}\,,
\end{equation}
where $\mu$ denotes the mass of the particle. The trajectories are given by
\begin{equation}\label{xdot0}
 {\dot x}^\mu(\tau)\,=\,{1\over \mu} {\partial{\cal H}\over \partial P_\mu} \,,
\end{equation}
and the dot denotes derivative with respect to the proper time, $\tau$. The Hamiltonian can be separated into a radial ${\cal H}_{r}$ and angular part ${\cal H}_{\theta}$,
 \begin{equation}
 {\cal H}+\frac{\mu^2}{2} = {1\over r^2} \left({\cal H}_{r} +  {\cal H}_{\theta}\right)\,.
\end{equation}
Thus, we can solve \eqref{hzero} by taking
\begin{equation}
\begin{aligned}
2 {\cal H}_{\theta} \,=\,&  -C^2+\frac{J^2}{\sin^2\theta}+P_{\theta}^2\,=\,0\,, \\[1mm]
2 {\cal H}_r\,=\,& C^2-\frac{r^2  E^2}{f N^2}+ r^2 f P_r^2+\mu ^2 r^2\,=\,0\,,
\end{aligned}
\end{equation}
where $J\,=\,P_\phi$, $E\,=\,-P_t$ and $C$ are constants of the motion.

A circular orbit along the equator is defined by the conditions
\begin{equation}
r\,=\,r_0\,, \hspace{1cm} \theta\,=\, \frac{\pi}{2}\,,
\end{equation}
 leading to $P_r(r_0)=P_r'(r_0)\,=\,0$ and $P_{\theta}\,=\,0$.
 Solving for $E$, $J$, and $C$, one finds
\be
 E\,=\,\frac{\mu F}{\sqrt{F - \ft{r}{2} F'}}  \Bigg|_{r=r_0} \,, \hspace{1cm} J \,=\,\frac{\mu r \sqrt{\ft{r}{2} F'} }{\sqrt{ F -  \ft{r}{2} F'}}  \Bigg|_{r=r_0} \,, \hspace{1cm} C\,=\, \pm J\,,
\ee
where
\begin{equation}\label{eq:defF}
F\,=\, f N^2\,.
\end{equation}
Equivalently, the geodesic can be parametrized in terms of two constants, $\Gamma$ and $\Omega_{\rm orb}$,
\begin{equation}\label{eq:circularorbits}
t(\tau)=\Gamma  \tau \, , \hspace{1cm} r(\tau)=r_0 \, , \hspace{1cm} \theta(\tau)=\ft{\pi}{2}\, , \hspace{1cm} \phi(\tau)= \Gamma \,\Omega_{\rm orb} \,\tau \, .
\end{equation}
These are related to $E$ and $J$ by
\begin{equation}
  \Gamma \,=\,    {E\over \mu F} \Bigg|_{r=r_0}\,, \hspace{1cm} \Omega_{\rm orb}= {J F  \over E r^2  }\Bigg|_{r=r_0} \label{gamma}\,.
\end{equation}
In terms of the velocity parameter,\footnote{We work in terms of this quantity for convenience, but let us note that this definition no longer coincides with the actual velocity of the orbit measured at infinity, $v\neq r_0 \Omega_{\rm orb}$.}
\begin{equation}
v^2\,=\,\frac{M}{r_0}\, ,
\end{equation}
one finds
\begin{equation}\label{eq:circular_orbits}
\begin{aligned}
E \,=\,& \frac{\mu  \left(1-2 v^2\right)}{\sqrt{1-3 v^2}}\left[1+\frac{\lambda_3}{M^4}\frac{10 v^{14} \left(-5+6 v^2\right)}{1-5 v^2+6 v^4}-\frac{\lambda_4}{M^6}\frac{128 v^{18} \left(14-42 v^2+33 v^4\right)}{1-5 v^2+6 v^4}\right] \,, \\[1mm]
 J \,=\,&   \frac{\mu  M}{v \sqrt{1-3 v^2}}\left[1-\frac{\lambda_3}{M^4}\frac{10 v^{12} \left(7-12 v^2\right)}{1-3 v^2}-\frac{\lambda_4}{M^6}\frac{64 v^{16} \left(36-119 v^2+99 v^4\right)}{1-3 v^2}\right]\,, \\[1mm]
\Omega_{\rm orb} \,=\,& \frac{v^3}{M} \left[1- \frac{70\lambda_3}{M^4}v^{12}+ \frac{64\lambda_4}{M^6}v^{16} \left(55 v^2-36\right) \right]\,, \\[1mm]
\Gamma \,=\,& \frac{1}{\sqrt{1-3v^2}}\left[1-\frac{\lambda_3}{M^4}\frac{90v^{14}}{1-3v^2}+ \frac{\lambda_4}{M^6}\frac{1408 v^{18} \left(3 v^2-2\right)}{1-3v^2}\right]\,.  \end{aligned}
\end{equation}
As we can see, the corrections to geodesic motion enter at 6PN ($v^{12}$) for the cubic corrections, and at 8PN for the quartic corrections controlled by $\lambda_{4}$.


\section{Gravitational perturbations }
\label{sec:grav_pert}
In this section we derive the master equations governing the dynamics of gravitational perturbations around the corrected Schwarzschild black hole presented in section~\ref{sec:correctedSchwarzschildsoln}. Namely, we consider a metric perturbation
\begin{equation}
g_{\mu\nu}\,=\, {\overline g}_{\mu\nu}+ \kappa^2 \, h_{\mu\nu}\,
\end{equation}
around the background metric ${\overline g}_{\mu\nu}$ given by \eqref{eq:bkgmetric}, and expand the metric field equations \eqref{eq:fieldeq} at linear order in the perturbation. We emphasize the fact that we are extending previous works \cite{Cardoso:2018ptl, Cano:2021myl,  deRham:2020ejn} by deriving the equations satisfied by the perturbations in the presence of an arbitrary source.

We shall proceed by decomposing the perturbation and source in spherical harmonics. As reviewed in appendix~\ref{app:spherical_harmonics}, these fall in two classes: odd and even harmonics, according to their behavior under parity. In a parity-preserving theory, such as the ones we are considering \eqref{eq:EFTofGR}, odd and even perturbations do not mix, and therefore we can analyze them separately. This is what we do next.

The linearized equations for the metric perturbation can be expressed as
\begin{equation}
    G_{\mu\nu}^{L}(h)=T^{\rm eff}_{\mu\nu}\, ,
\end{equation}
where $G_{\mu\nu}^{L}(h)$ is the linearized Einstein tensor evaluated on $h_{\mu\nu}$.
The main difference with respect to GR is that now we deal with an effective stress-energy tensor,
\begin{equation}\label{eq:Tmunueff}
T^{\rm eff}_{\mu\nu}\,=\,  T_{\mu\nu}-{\cal E}^{L}_{\mu\nu}(h)\, ,
\end{equation}
accounting for the contributions of the actual source and of the higher-derivative interactions. The latter are encoded in ${\cal E}^{L}_{\mu\nu}(h)$, which is defined as the linearized tensor appearing in \eqref{eq:Emunu}. Although this contains higher-order derivatives of the perturbation $h_{\mu\nu}$, we can make use of the zeroth-order equations satisfied by $h_{\mu\nu}$ in order to replace them by $h_{\mu\nu}$ itself and its first derivative.

\subsection{Odd perturbations}

Our analysis of odd perturbations follows the original work by Regge and Wheeler \cite{Regge:1957td} to a large extent. To express the metric perturbation, it is convenient to introduce a label for the coordinates on the sphere, $x^{A}=\{\theta, \phi\}$. Then, we have that
\begin{equation}\label{hodd}
h_{\mu\nu}^{\rm odd} \diff x^{\mu} \diff x^{\nu} \,=\,  e^{-{\rm i} \omega t} \left[2\left(h_0(r) \diff t + h_1(r) \diff r\right) X^{\ell m}_A  \diff x^A+ h_2(r) X^{\ell m}_{AB}\, \diff x^A \diff x^B\right]\,,
\end{equation}
where $X_{A}^{\ell m}$ and $X_{AB}^{\ell m}$ are the odd vector and tensor spherical harmonics, defined by \eqref{eq:oddvectorharmonics} and by \eqref{eq:oddtensorharmonics}. From now on we shall set ourselves in the Regge-Wheeler gauge \cite{Regge:1957td}, which implies
\be
h_2(r)\,=\,0\,.
\ee
Likewise, the source $T_{\mu\nu}$ and the effective stress-energy tensor \eqref{eq:Tmunueff} can be decomposed in odd spherical harmonics,
\begin{equation}\label{todd}
T_{\mu\nu}^{\rm odd} \diff x^{\mu} \diff x^{\nu} \,=\,  e^{-{\rm i} \omega t} \left[2\left({\cal S}_0(r) \diff t + {\cal S}_1(r) \diff r\right) X^{\ell m}_A  \diff x^A+ {\cal S}_2(r) X^{\ell m}_{AB}\, \diff x^A \diff x^B\right]\,,\\[1mm]
\end{equation}
\begin{equation}
T_{\mu\nu}^{\rm {eff}, \rm{odd}} \diff x^{\mu} \diff x^{\nu} \,=\,  e^{-{\rm i} \omega t} \left[2\left({\cal S}^{\rm eff}_0(r) \diff t + {\cal S}^{\rm eff}_1(r) \diff r\right) X^{\ell m}_A  \diff x^A+ {\cal S}^{\rm eff}_2(r) X^{\ell m}_{AB}\, \diff x^A \diff x^B\right]\,.
\end{equation}
Plugging \eqref{hodd} and \eqref{todd} into \eqref{eq:fieldeq}, one finds the following system of two first-order differential equations for $h_0$ and $h_1$:\footnote{One can check that the remaining equations are implied by \eqref{einst13} and \eqref{einst23} provided we impose $\nabla_{\mu}T^{\mu\nu}\,=\,0$.}
\begin{align}
\label{einst13}
  &&-2  r^2 N {\cal S}_0^{\text{eff}}+  r^2 f \left(-N h_0''+N' h_0'\right) -\ii \omega  \,r^2 f N h_1'+\ii \omega \, r f \left(-2N+r N'\right)h_1 \nn\\[1mm]
   &&  +h_0\left[N\left(\nu+2f+\left(r^2 f'\right)'\right)+r^2\left(3f' N'+2 f N''\right)\right] \,=\,0\,, \\[2mm]
    \label{einst23}
   &&2 r^2 f N^2
   {\cal S}_1^{\text{eff}} + \ii \omega \left(-r^2
   h_0'+2 r h_0 \right)+h_1\left[-f N\left(3r^2f'N'+N\left(\nu+2f+\left(r^2 f'\right)'\right)\right)\right.\nn\\[1mm]
   &&\left.+\omega^2r^2-2rf^2 N\left(rN'\right)'\right] \,=\,0\,,
\end{align}
where we have defined
\be
\nu\,=\,\left(\ell+2\right) \left(\ell-1\right)\,.
\ee
 The above system is equivalent to a second-order differential equation for a single variable. For our purposes here, it is convenient to choose the following combination:
\be\label{psicpm}
\Psi_{\rm odd}(r)\,=\,\frac{2 r A_{\rm odd}(r)}{\nu} \left[h_0'+\ii\omega h_1 -\frac{2}{r} \,h_0\right]\,,
\ee
which coincides with the master variable introduced by Cunningham, Price and Moncrief in \cite{Cunningham:1978zfa}, up to the global factor $A_{\rm odd}$.  This is fixed to
\begin{equation}
A_{\rm odd}(r)\,=\, 1+ \frac{156\lambda_3M^2}{r^6}+1152\lambda_4 M^2\left(-\frac{167M}{9r^9}+\frac{9}{r^8}\right)+\frac{1152 {\tilde\lambda}_4M^2 \left(\nu+2\right)}{r^8}\,
\end{equation}
by requiring that the master equation satisfied by \eqref{psicpm} takes the following form
\be\label{eq:mastereqodd}
\frac{\diff}{\diff r}\left[f N \frac{\diff \Psi_{\rm odd}}{\diff r}\right] + \frac{1}{fN}\left(\omega^2 - f V_{\rm odd} \right)\Psi_{\rm odd}\,=\, S_{\rm odd}\, ,
\ee
where
\begin{equation}
\begin{aligned} \label{vodd}
V_{\rm odd}\, =& \, \frac{\nu+2}{r^2}-\frac{6 M}{r^3} +  V^{(3)}_{\rm odd} + V^{(4)}_{\rm odd} \,, \\[1mm]
S_{\rm odd}\,=& \,\frac{-4 r {\cal S}_0'-4 \ii \omega r  {\cal S}_1}{\nu}+  S^{(3)}_{\rm odd} + S^{(4)}_{\rm odd} \,.
\end{aligned}
\end{equation}
The corrections to the potential $V_{\rm odd}$ and the source term $S_{\rm odd}$ for each of the corrections are summarized below.

\paragraph{Cubic corrections:}

\begin{eqnarray}
V^{(3)}_{\rm odd}&\,=\,&-\frac{72 M^2 \lambda_3}{r^8} \left[ \left(5\nu-81\right) +\frac{173 M}{r}\right]+\frac{864 \lambda_3 M \omega ^2}{r^5}\,, \label{vodd3}\\[1mm]
 S^{(3)}_{\rm odd}&\,=\, & \frac{48 M \lambda_3 }{\nu\left(2 M-r\right)r^6 }\left\{\left[\left(2 M-r\right) \left(5 \left(\nu-10\right) r+216 M\right)-7 r^4 \omega ^2\right]{\cal S}_0\right.\\[1mm]
 &&+\ii \omega r \left[\left(r-2 M\right) \left(\left(\nu-52\right) r+138 M\right)-r^4 \omega
   ^2\right]{\cal S}_1\nn\\[1mm]
   &&+\left[r \left(r-2 M\right) \left(\left(\nu-34\right) r+88 M\right)-r^5 \omega ^2\right]{\cal S}_0'+2 \ii \omega r^2 \left(r-2 M\right) \left(8 r-17 M\right){\cal S}_1'\nn\\[1mm]
   &&\left.r^2 \left(20 M-9 r\right) \left(2 M-r\right) {\cal S}''_0 -\ii r^3 \omega  \left(r-2 M\right)^2 {\cal S}''_1-r^3 \left(r-2 M\right)^2{\cal S}'''_0\right\}\,.\nn \label{sodd3}
\end{eqnarray}

\paragraph{Quartic corrections:}

\begin{eqnarray}\label{vodd4}
V^{(4)}_{\rm odd}&\,=\,&\frac{256 M^2 \lambda_4}{r^{10}} \left[\frac{(139 \nu-13231) M}{r} +36 \left(77-2 \nu\right)+\frac{15603 M^2}{r^2}\right]\\[1mm]
&&+\frac{16128 M^2\lambda_4 \omega^2}{r^8}-\frac{1152 \nu \tilde{\lambda_4} M^2 \left(\nu+2\right)   }{r^{10}} \nn\,,\\[1mm]
S^{(4)}_{\rm odd}&\,=\,&\frac{1536 M^2 \lambda_4 }{\nu \left(2 M-r\right)r^9}\left\{\left[2 \left(2 M-r\right) \left(45 M-4 \left(\nu+2\right) r\right)-8 r^4 \omega ^2\right]{\cal S}_0\right.\\[1mm]
&&+ \ii \omega r \left[r^4 \omega^2 -\left(r-2 M\right) \left(\left(\nu-49\right) r+103 M\right)\right]{\cal S}_1\nn \\[1mm]
&&+\left[r^5 \omega ^2-r \left(r-2 M\right) \left(\left(\nu-103\right) r+227 M\right)\right]{\cal S}'_0-2 \ii\omega r^2 \left(11 M-5 r\right) \left(2 M-r\right){\cal S}'_1\nn\\[1mm]
&&\left. +r^2 \left(2 M-r\right) \left(18 r-38 M\right){\cal S}''_0+\ii \omega r^3 \left(r-2 M\right)^2 {\cal S}''_1+r^3 \left(r-2 M\right)^2{\cal S}'''_0\right\}\,.\nn
\end{eqnarray}

\subsection{Even perturbations}

Let us now consider the even perturbations. The decomposition of $h_{\mu\nu}$ into spherical harmonics of even type is as follows:
\begin{equation}\label{heven}
\begin{aligned}
h_{\mu\nu}^{\ev} \diff x^\mu \diff x^\nu \,=\, &  e^{-{\rm i} \omega t} \left[\left(H_{00} \,\diff t^2 +H_{11} \,\diff r^2+2 H_{01} \,\diff t \,\diff r\right)Y^{\ell m}  \right. \\[1mm]
& \left. +2\left(j_0 \,\diff t + j_1 \,\diff r\right)Y^{\ell m}_{A} dx^{A}+ r^2\left(G Y^{\ell m}_{AB}+ K \Omega_{AB}Y^{\ell m}\right)\diff x^A \diff x^B\right]  \,,
\end{aligned}
\end{equation}
where $H_{00}, H_{01}, H_{11}, j_0, j_1, G$ and $K$ depend only on $r$, $Y^{\ell m}_{A}$ and $Y^{\ell m}_{AB}$ are defined in \eqref{eq:evenvectorharmonics} and \eqref{eq:eventensorharmonics}, and $\Omega_{AB}$ is the metric of the unit two-sphere. Following \cite{Zerilli:1974ai}, we use the gauge freedom to fix
 \be
 j_0\,=\,j_1\,=\,G\,=\,0\, .
 \ee
Next, we expand the source $T_{\mu\nu}$ in spherical harmonics of even type as well:
\begin{equation}\label{teven}
\begin{aligned}
T_{\mu\nu}^{\ev} \diff x^\mu \diff x^\nu \,=\, &  e^{-{\rm i} \omega t} \left[\left(\T_{00} \,\diff t^2 +\T_{11} \,\diff r^2+2 \T_{01} \,\diff t \,\diff r\right)Y^{\ell m}  \right. \\[1mm]
& \left. +2\left(\T_0 \,\diff t + \T_1 \,\diff r\right)Y^{\ell m}_{A} dx^{A}+\left({\cal Q} \,Y^{\ell m}_{AB}+ {\cal P} \,\Omega_{AB}Y^{\ell m}\right)\diff x^A \diff x^B\right]  \,.
\end{aligned}
\end{equation}
As it occurs in GR, when substituting these ansatz\"e on the linearized field equations one finds that they amount to two algebraic constraints between the variables $H_{00}, H_{11}, H_{01}, K$, and a coupled system of two first-order diferential equations. Solving the algebraic constraints for $H_{11}$ and $H_{00}$, for instance, we are left with a coupled system of two first-order ODEs for the variables $K$ and  $H_{01}$. This is equivalent to a single second-order differential equation, which can be conveniently written by introducing the even master variable
\begin{equation}
\Psi_{\ev}(r)\,=\, \frac{2r D_{\ev}(r)}{\nu+\frac{6M}{r}}\left(K+\frac{f}{r}\frac{H_{01}}{\ii \omega}\right)\,,
\end{equation}
where
\begin{equation}
\begin{aligned}
 D_{\ev}(r)\,=\,&1+\frac{24 \lambda_3 M}{6 M+\nu r} \left[\frac{-63 M^2}{r^6}-\frac{2 \left(8 \nu -9\right) M}{r^5}+\frac{6\nu}{r^4}\right]\\[1mm]
 &+\frac{384 \lambda_4 M^2}{6 M+\nu r} \left[-\frac{82 M^2}{r^9}+\frac{3 M \left(\nu+16\right)}{r^8}+\frac{\nu \left(\nu+2\right)}{r^7}\right]\,.
\end{aligned}
\end{equation}
The differential equation satisfied by $\Psi_{\ev}(r)$ is
\begin{equation}
\label{eq:mastereven}
\frac{\diff}{\diff r}\left[N f \frac{\diff\Psi_{\ev}}{\diff r}\right]
+\frac{1}{N f}\left(\omega^2 - f V_{\ev}\right)\Psi_{\ev}\,=\, {S}_{\ev}\, ,
\end{equation}
where
\begin{equation}\label{veven}
\begin{aligned}
 V_{\ev}\, &=\,\frac{72 M^3+36 M^2 \nu r+6 M \nu^2 r^2 +r^3\nu^2\left(\nu+2 \right)}{r^3 \left(6 M+\nu r\right)^2}+  V^{(3)}_{\ev} + V^{(4)}_{\ev}\,, \\[1mm]
{S}_{\ev}&\,=\,-\frac{2 {\cal Q}}{r}-\frac{2 (2 M-r) \left[r \omega  \T_{11}+2 \omega  \T_1-\ii \left(r \T_{01}'+2 \T_0'\right)\right]}{(6 M+L r)\omega }+S^{(3)}_{\ev}+S^{(4)}_{\ev} \\[1mm]
  &+\frac{2\ii}{\omega }\,\frac{2 M r \left[6 M-r \left(\nu+6\right)\right] \T_{01}+\left[24 M^2+2 M r \left(\nu-6\right)+r^2 \nu\left(\nu+2\right)\right] \T_0}{r\left(6 M+\nu r\right)^2 } \,. \\[1mm]
\end{aligned}
\end{equation}
The corrections to $V_{\ev}$ from the curvature invariants controlled by $\lambda_3$ and $\lambda_4$ are displayed below. Those to $S_{\ev}$ are too long, so we retain ourselves from presenting them in the main text. The reader can find them in the Mathematica file included in the arXiv submission.

\paragraph{Cubic corrections:}

\begin{equation} \label{veven3}
\begin{aligned}
V^{(3)}_{\ev}\,=\,&\frac{72M \lambda_3 }{\left(6M+\nu r\right)^3r^9}\left[23784 M^5+28 \left(-378+757 \nu\right) M^4 r\right.\\[1mm]
&+6 \nu \left(-1842+1055 \nu\right) M^3 r^2+\nu \left(720-3598 \nu+717 \nu^2\right) M^2 r^3\\[1mm]
&\left.+3 \nu^2 \left(112-139 \nu+7 \nu^2\right) M r^4-10
   \left(-4+\nu\right) \nu^3 r^5\right]-\frac{288 M\lambda_3\omega^2}{r^5}\, .
\end{aligned}
\end{equation}

\paragraph{Quartic corrections:}

\begin{equation} \label{veven4}
\begin{aligned}
V^{(4)}_{\ev}&\,=\,
\frac{128M^2 \lambda_4 }{\left(6M+\nu r\right)^3r^{12}}\left[236736 M^5 + 144 \left(-675 + 2012 \nu\right) M^4 r\right.\\[1mm]
&+324 \nu \left(-626+253 \nu\right) M^3 r^2+6 \nu \left(6048+\nu \left(-11066+991 \nu\right)\right) M^2 r^3 \\[1mm]
&\left.-2 \nu^2 \left(-7560+\nu \left(2591+58 \nu\right)\right) M r^4+9 \nu^3 \left(176+\nu \left(18+\nu\right)\right) r^5\right]+\frac{16128 M^2\lambda_4\omega^2}{r^8}\,.
\end{aligned}
\end{equation}
Let us notice that the term controlled by ${\tilde \lambda}_{4}$ does not modify the equations for even perturbations.


\section{Scalar waveform}\label{sec:scalar}

In order to illustrate our calculations in a simplified setting, let us consider first the scalar-wave emission by a charged point-like source on circular motion around the black hole geometry \eqref{eq:bkgmetric}.

\subsection{Scalar-wave equation}

For simplicity, let us take a scalar field $\varphi$ minimally coupled to the metric. Concretely, we modify our effective action \eqref{eq:EFTofGR} as follows
\begin{equation}
S_{\rm EFT}  \to S_{\rm EFT} -\int \diff^4 x\sqrt{-g}\left[ \frac{1}{2}\,\partial_{\mu}\varphi \,\partial^{\mu}\varphi +  \left(1+\alpha\varphi\right) \varrho\right]\,,
\end{equation}
where $\alpha$ is a coupling constant and
\be
\varrho\,=\, \mu  \int  {\diff \tau\over \sqrt{-g}}\,  \delta^{(4)}\left(x-\bar{x}(\tau)\right)  \,
\ee
is the Lagrangian density associated to a massive  particle of mass $\mu$.
The scalar equation of motion  reads
\begin{equation}
\partial_{\mu}\left(\sqrt{-g}\,\partial^\mu\varphi\right)\,=\, \sqrt{-g}\, \alpha\varrho\, .
\end{equation}
Expanding in spherical harmonics both the perturbation and the source,\footnote{See appendix~\ref{app:spherical_harmonics} for our conventions.}
\begin{equation}
\begin{aligned}
\varphi\,=\,& \int {\diff \omega \over 2\pi}  \sum_{\ell m} e^{ -\ii \omega t} \varphi_{\ell m }(r) Y^{\ell m}\left(\theta, \phi\right)\,,\\[1mm]
\sqrt{-g} \,\alpha\varrho\,=\,& \int {\diff \omega\over 2\pi}  \sum_{\ell m} e^{ -\ii \omega t} S_{\ell m }(r) Y^{\ell m}\left(\theta, \phi\right)\,,
\end{aligned}
\end{equation}
we end up with the following differential equation for $\varphi_{\ell m }$:
\begin{equation}
\label{eq:scalarwave}
 \frac{\diff}{\diff r}\left[r^2 f N \frac{\diff\varphi_{\ell m }}{\diff r}\right]+r^2 N\left(\frac{\omega^2}{f N^2}-\frac{\ell \left(\ell+1\right)}{r^2}\right)\varphi_{\ell m }\,=\, S_{\ell m }\, .
\end{equation}
For a charged test particle on circular motion, we can make use of \eqref{eq:circularorbits} and of the identity\footnote{Bar denotes complex conjugation.}
\begin{equation}
\delta\left(\theta-\frac{\pi}{2}\right)\delta\left(\phi-\phi_0\right)\,=\, \sum_{\ell m}Y^{\ell m}(\theta, \phi) {\bar Y}^{\ell m}\left(\frac{\pi}{2},\phi_0\right)\,,
\end{equation}
to find that the source term is given by
\begin{equation}\label{eq:Slmpp}
S_{\ell m}\,= \,\frac{2\pi \mu \alpha}{\Gamma}\,\kappa^{\ell m}_0\delta\left(r-r_0\right)\delta\left(\omega-m\Omega_{\rm orb}\right)\,,
\end{equation}
where we have defined
\begin{equation}
\kappa^{\ell m}_0\,=\,{\bar Y}^{\ell m}\left(\ft{\pi}{2}, 0\right) \,.
\end{equation}
The solution of \eqref{eq:scalarwave} satisfying appropriate boundary conditions at the horizon $(r=r_+)$ and at infinity ($r\to \infty$) is given by
 \be\label{rt}
\varphi_{\ell m}(r)\,=\,  \int_{r_+}^\infty G( r ,  r') S_{\ell  m}(r') \diff r' \,,
\ee
where
\be
G(r,r')\,=\,
{1\over W}   \left\{
\begin{array}{ccc}
  \varphi_{\rm in}( r')\,\varphi_{\rm out} ( r)  & ~~~~~~~&  r'< r \\[1mm]
 \varphi_{\rm in} (r)\,\varphi_{\rm out} ( r')  & ~~~~~~~ &   r< r' \\
\end{array}
\right.\label{green}
\ee
is the Green function built out of two independent solutions of the homogenous equation. These satisfy incoming ($\varphi_{\rm in}$) and outgoing ($\varphi_{\rm out}$) boundary conditions at the horizon and infinity, respectively. More conceretely, we require the following behavior at infinity,
\begin{equation}\label{eq:bdrycondinf}
\begin{aligned}
  \varphi_{\rm in}(r) \,\underset{r\to \infty}{\longrightarrow}\,&  B^{\rm in}_{+}  \,   \frac{e^{{\rm i} \omega r_*}}{r} + B^{\rm in}_{-} \,  \frac{e^{-{\rm i} \omega r_*}}{r} \,, \\[1mm]
 \varphi_{\rm out}(r) \underset{r\to \infty}{\longrightarrow}&   B^{\rm out}_{+}  \,   \frac{e^{{\rm i} \omega r_*}}{r}  \,,
\end{aligned}
\end{equation}
where $r_*$ is the GR tortoise coordinate,
\begin{equation}\label{eq:tortoise}
r_*\,=\,r+2M\log \left(\frac{r}{2M}-1\right)\,.
\end{equation}
Instead, at the horizon we impose,
\begin{equation}\label{eq:bdrycondhor}
\begin{aligned}
  \varphi_{\rm in}(r) \,\underset{r\to r_+}{\longrightarrow}\,&  D^{\rm in}_{-}  \,\left(1-\frac{r_+}{r}\right)^{- \frac{\ii\omega}{4\pi T}}  \,,\\[1mm]
 \varphi_{\rm out}(r) \,\underset{r\to r_+}{\longrightarrow}\,&  D^{\rm out}_{-}  \,\left(1-\frac{r_+}{r}\right)^{-\frac{\ii\omega}{4\pi T}} +D^{\rm out}_{+}  \,\left(1-\frac{r_+}{r}\right)^{\frac{\ii\omega}{4\pi T}} \,,
\end{aligned}
\end{equation}
where $T$ is the temperature of the black hole given in \eqref{eq:inversetemperature}.
Finally, $W$ is a constant built out of the Wronskian,
\be\label{ww}
W\,=\, r^2 N f   \left(\varphi_{\rm in}   \frac{\diff\varphi_{\rm out}}{\diff r} -  \varphi_{\rm out} \frac{\diff\varphi_{\rm in}}{\diff r} \right)   \,.
\ee
The constancy of $W$ ---which follows from the fact that $\varphi_{\rm in}$ and $\varphi_{\rm out}$ are solutions of the homogeneous part of (\ref{eq:scalarwave})--- allows us to compute it at infinity, where we can make use of \eqref{eq:bdrycondinf} to obtain that
\begin{equation}
W\,=\, 2\ii \omega B^{\rm out}_{+}B^{\rm in}_{-}\, .
\end{equation}
In the following, we normalize the incoming and outgoing solutions imposing that
\begin{equation}\label{binomega}
B_-^{\inn}\,=\, \frac{1}{2\ii\omega}\,, \hspace{1cm} B_+^{\out}\,=\,1\,,
\end{equation}
so that $W=1$.  Far away from the source, the solution to the inhomogeneous equation \eqref{rt} can be approximated by
\be\label{phiints}
\begin{aligned}
\varphi_{\ell m}(r) \underset{r\to \infty}{\longrightarrow}&  \frac{e^{\ii\omega r_*}}{r}  Z_{\ell m}\,,
\end{aligned}
 \ee
 where the (scalar) waveform $Z_{\ell m}$ is given by
\begin{equation}
Z_{\ell m}\,=\,\int_{r_+}^\infty  \varphi_{\rm in}(r')    {S}_{\ell  m}(r') \,\diff r'=\tilde{Z}
_{\ell m} \delta\left(\omega-m\Omega_{\rm orb}\right) \, .
\end{equation}
For a point-like source in circular motion, this integral can be evaluated straightforwardly using \eqref{eq:Slmpp}, yielding
\begin{equation}\label{eq:scalarwaveform}
\tilde{Z}_{\ell m}\,=\,\frac{2\pi \mu  \alpha}{\Gamma}\,\kappa^{\ell m}_0 \varphi_{\rm in}(r_0) \, .
\end{equation}
Therefore, all we need to extract ${\tilde Z}_{\ell m}$ is to find a solution of the homogeneous equation satisfying incoming boundary conditions at the horizon. This task is now addressed.

\subsection{The incoming solution}

Let us consider the homogeneous part of the scalar-wave equation \eqref{eq:scalarwave},
\begin{equation}\label{boxlm0hom}
  \frac{\diff }{\diff  r}\left(r^2 f N \frac{\diff\varphi_{\ell m}  }{\diff r}\right) +r^2 N  \left(\frac{\omega^2}{f N^2}-\frac{\ell \left(\ell+1\right)}{r^2}\right) \varphi_{\ell m} \,=\,0\,.
\end{equation}
Next, we introduce the dimensionless quantities,
\begin{equation}\label{eq:definitions}
z\,=\, \frac{2M}{r}\,, \hspace{1cm} x\,=\, 4\ii M\omega\,, \hspace{1cm} y\,=\, \frac{x}{z}\,=\, 2\ii\omega r\, ,
\end{equation}
and look for solutions of (\ref{boxlm0hom}) in the regime where $x, z$ are small but their ratio $y=\frac{x}{z}$ is finite. We will refer to this limit as the PM expansion. Performing the following change of variable,
\begin{equation}\label{phig}
\varphi_{\ell m}\left(\frac{2M y}{x}\right)\,=\,e^{-\frac{y}{2}} (1-\ft{x}{y})^{-{x\over 2} }  y^{-1-\frac{x}{2}}  G(y)  \,,
\end{equation}
our original equation \eqref{boxlm0hom} can be cast in the form
 \be\label{heundef}
\left({\cal L}_{\rm CHE} + {\cal L}_{\lambda} \right) G\,=\,0  \,,
\ee
where ${\cal L}_{\rm CHE}$ is the confluent Heun operator
\begin{equation}\label{lCHE}
\begin{aligned}
{\cal L}_{\rm CHE}\, G\,&=\, \frac{\diff^2 G}{\diff y^2}+\frac{x (-1+m_1+m_2+2 m_3-y)+y (-2 m_3+y)}{(x-y) y}\frac{\diff G}{\diff y}\\[1mm]
&+ \frac{4 \left(-1+m_1+m_3\right) \left(-1+m_2+m_3\right) x+\left(-1+4 m_3-4 m_3^2+4 u\right) y}{4 (x-y)
   y^2}G\,.
   \end{aligned}
   \end{equation}
The parameters of the specific CHE arising are related to $x$ and $\ell$ by
\begin{equation}
    m_1 \,=\, m_2\,=\,-m_3\,=\,  \frac{x}{2}\,,   \hspace{1cm} u\,=\, \left( \ell+{1\over 2} \right)^2+ \frac{x^2}{4}+\frac{x}{2}  \,.
\end{equation}
In turn, ${\cal L}_{\lambda}$ contains all terms linear in the higher-derivative couplings. We avoid presenting the explicit expression for ${\cal L}_{\lambda}$ here as it is considerably long and not particularly illuminating.

Thus, we have that  \eqref{heundef} reduces to a CHE for $\lambda_3=\lambda_4=0$.\footnote{Note that ${\tilde \lambda}_4$ does not appear in \eqref{heundef} as it does not modify the black hole geometry in the static case.}  Following the approach of \cite{Cipriani:2025ikx}, two independent solutions of the homogoneous equation can be found recursively, order by order in $x$, in terms of a single hypergeometric function.
 Here we show that the same procedure works for the corrected Heun equation as well.
 More precisely, we find that two independent solutions of (\ref{heundef}) can be written as
\be \label{collection1}
G_\alpha (y) \,=\, P(y) \, H_\alpha(y) + \widehat{P}(y) \,  y  \frac{\diff H_\alpha{}(y)}{\diff y} \,, \hspace{1cm} \alpha\,=\, \{+1, -1\} \,,
\ee
where
 \be \label{h012}
H_\alpha(y) \,=\, y^{\frac{1}{2}-\alpha a-m_3} \,
   _1F_1\left(\ft{1}{2}-\alpha a-m_3;1-2\alpha
   a;y\right)\,,
\ee
and
 \begin{equation}
\begin{aligned}
P(y) \,=\,& 1+\sum_{i=1}^k   \sum_{j=0}^{i-1}   c_{ij} \, x^i y^{j-i} \,,  \hspace{1cm} \widehat{P}(y)\,=\, \sum_{i=1}^k   \sum_{j=0}^{i-1}   \widehat{c}_{ij} \, x^i y^{j-i} \,.
\end{aligned}
\end{equation}
 The parameter $a$ is related to $u$ (that is, $\ell$) via
 \be\label{ua}
 u\,=\,\left( \ell+{1\over 2} \right)^2+ \frac{x^2}{4}+\frac{x}{2}\,=\, a^2+\sum_{i=1}^\infty u_i x^i \, .
 \ee
Plugging this ansatz in \eqref{heundef}, one finds a set of algebraic equations for the coefficients $c_{ij}, {\widehat c}_{ij}$ and $u_i$. The higher-derivative couplings $\lambda_3$ and $\lambda_4$ modify these coefficients, and therefore the homogeneous solutions. In order to present our results, it is convenient to introduce some notation for the GR and higher-derivative contributions,
\be
P(y)\,=\,P_{{\rm GR}}(y)+\delta P(y) \,, \hspace{1cm} \widehat{P}(y)\,=\,\widehat{P}_{{\rm GR}}(y)+\delta \widehat{P}(y)\,, \hspace{1cm} a\,=\, a_{\rm GR}+ \delta a\,.
\ee
Moreover, bearing in mind that later we will have to evaluate the solution at the position of the source, we will organize the expressions in a PN expansion ---namely, a expansion for small $v$---, taking into account that
\begin{equation}\label{eq:PNregime}
 z_0\equiv \frac{2M}{r_0}\,=\, 2v^2\,,  \hspace{1cm} x_0\equiv 4\ii m M \Omega_{\rm orb}\sim v^3\, , \hspace{1cm} y_0\equiv \frac{x_0}{z_0}\sim v\,,
\end{equation}
so that a term $x_0^n y_0^m \sim v^{3n+m}$ will produce a contribution at $\tfrac{3n+m}{2}$PN order. This being said, we have that the GR contribution up to 2.5PN is given by
\begin{equation}\label{plambda}
\begin{aligned}
 P_{{\rm GR}}(y) &\,=\,1+\frac{x }{2 y} +\frac{\left(16 a_{\rm GR}^4 +24 a_{\rm GR}^2- 39\right) x^2 }{128 \left(a_{\rm GR}^2-1\right) y^2} +\frac{\left(27-28 a_{\rm GR}^2\right) x^2 }{128
   \left(a_{\rm GR}^2-1\right) y}   +\ldots\,, \\[1mm]
      \widehat{P}_{{\rm GR}}(y) &\,=\,-\frac{x }{2 y}  +\frac{\left(21-20 a_{\rm GR}^2\right) x^2 }{64 \left(-1+a_{\rm GR}^2\right) y^2}  +\frac{x^2 }{8 y}  +\ldots\,.
\end{aligned}
\end{equation}
Just to show the leading effect from the higher-derivative terms, we display the expressions for the polynomials $\delta P(y)$, ${\delta \widehat P(y)}$ up to 6PN order,
\begin{equation}
\delta P(y)  \,=\,\frac{9 \left(61-4 a_{\rm GR}^2\right) x^6  \lambda _3}{256 \left(-9+a_{\rm GR}^2\right) M^4 y^6}   +\ldots \,, \hspace{1cm}
\delta\widehat{P}(y)  \,=\,   \frac{9 \left(-31+4 a_{\rm GR}^2\right) x^6  \lambda _3}{128 \left(-9+a_{\rm GR}^2\right) M^4 y^6} + \ldots \,,
\end{equation}
even though we will not make use of them in this section. In the above formulas the $\lambda_4$-corrections do not appear as they enter at 8PN. Finally, we provide the expression for $a\,=\, a_{\rm GR}+\delta a$ in terms of $\ell$:
\be\label{al}
\begin{aligned}
a_{\rm GR}\,=\,&\ell+\ft12 + \frac{\left(15 \ell^2+15 \ell-11\right) x^2}{8 (2 \ell+1) \left(4 \ell^2+4 \ell-3\right)}+{\cal O}(x^4) \,,\\[1mm]
\delta a\,=\,& -\frac{27 \lambda_3\left(-15+\ell+\ell^2\right) x^6}{128M^4 (1+2 \ell) \left(-35+4 \ell+4 \ell^2\right) \left(-15+4 \ell+4 \ell^2\right) \left(-3+4 \ell+4
   \ell^2\right)} + {\cal O}(x^8)\, .
\end{aligned}
\ee
The complete expression for the polynomials and the parameter $a$ up to order $x^6$ can be found in the ancillary Mathematica file.

Given the two independent solutions $G_\pm(y)$, one can build  any solution of the homogeneous equation as a linear combination. In particular, for the incoming solution we write
\begin{equation}\label{eq:incomingsoln_scalar}
\varphi_{\rm in}\left(\frac{2M y}{x}\right)  \,=  C_{\rm in} \,e^{-\frac{y}{2}} \left(1-\frac{x}{y}\right)^{-{x\over 2} } y^{-1-\frac{x}{2}}   \left[G_{-}\left(y\right)+ x^{2a}  F_\ell  \,G_{+}\left(y\right) \right] \,,\\[1mm]
\end{equation}
 where $F_{\ell}$ is a constant to be determined by imposing incoming boundary conditions at the horizon and $C_{\rm in} $ is an overall normalization. The behavior of the incoming solution \eqref{eq:incomingsoln_scalar} when $y\to \infty$ is obtained by using the following property of the hypergeometric functions,
\begin{equation}
H_\alpha(y)  \underset{y\to \infty}{\longrightarrow}   \sum_{\beta=\{+, -\} } B_{\alpha \beta }  \,  y^{-m_3(1+\beta ) }
e^{ y (1+\beta)\over 2}\,,
\label{eq:conform}
\end{equation}
where
\be \label{bab}
B_{\alpha \beta} =   e^{\frac{{\rm i} \pi \left(1-\beta\right)}{2}   \left({1\over 2}-m_3-\alpha a\right)}  \frac{\Gamma (1-2  \alpha a )   }{\Gamma \left(\ft{1}{2}-\alpha a- \beta   m_3  \right)}\,, \hspace{1cm} \alpha, \beta\,=\, \{+,-\}\, .
\ee
Moreover, in this limit we have that $P(y)\underset{y \to \infty}{\longrightarrow} 1$, which leads to
\begin{equation}\label{inf}
\varphi_{\rm in}(r)  \,\underset{r\to \infty}{\longrightarrow}  \frac{C_{\rm in}}{2\ii\omega}   \,{ e^{-{\rm i} \omega  r_* }  \over r}  x^{-\frac{x}{2}}    \left(B_{--}+ x^{2a}  F_\ell  \,B_{+-}\right)  +B_+^{\out}  {e^{{\rm i} \omega r_*}  \over r} \,.\\[1mm]
\end{equation}
Imposing the normalization \eqref{binomega}, one finds
 \be\label{cin}
 C_{\rm in} \,=\, {x^{x\over 2} \over B_{--}  \left(1 + x^{2a} F_\ell \ft{B_{+-}}{B_{--} }\right)}  \, .
 \ee

 \subsection{Love numbers}\label{sec:scalarLove}

The last ingredient we need is the coefficient $F_\ell$ specifying the linear combination entering in the incoming solution (\ref{eq:incomingsoln_scalar}). In the \emph{near-zone}, $x\ll z \ll 1$, where distances are much larger than the Schwarzschild radius but still smaller than the wavelength, the incoming solution behaves as
\begin{equation}\label{eq:incomingsoln_scalar2}
\varphi_{\rm in}\left(\frac{2M}{z}\right)  \,\underset{x\ll z \ll 1}{=}   C_{\rm in} x^{\hat{\ell}}  z^{-\hat{\ell}} \left(1+\ldots\right) \left[1 + F_{\ell} \,z^{2\hat{\ell} + 1} \left(1 + \ldots \right)  \right]  \,,\\[1mm]
\end{equation}
 where the dots stand for polynomial terms in $z$ and $\hat{\ell}=a-\ft12=\ell+O( \omega^2 M^2)$. As it turns out, $F_{\ell}$ computes the Love number of the geometry ---namely, the ratio between the coefficient of the term $z^{\hat{\ell}+1} $ describing the response of the geometry and that of the source term $z^{-\hat{\ell}}$---, and it is determined by imposing incoming boundary conditions at the horizon. In GR, a solution that covers the region close to the horizon can be obtained by using a different hypergeometric ansatz for $G(z)$, expanding again for small $x$ but keeping now $z$ finite \cite{Cipriani:2025ikx}. Unfortunately, such ansatz does not work when including the higher-derivative corrections. Still, as we show now, the corrected scalar-wave equation \eqref{eq:scalarwave} can be solved perturbatively in $x$ once we specify the value of $\ell$.\footnote{At least for the first $\ell$-modes, and up to certain order in $x$.} In this section, we will compute the scalar waveform and energy flux at 4PN, so as to account for the leading and next-to-leading contributions from the higher-derivative interactions. To this aim, all we need is to find the incoming solution for the $\ell\,=\,1$ mode in the static $x\to 0$ limit. The reason is that tidal interactions enter in the waveform at $(2\ell+1)$PN order, so only the $\ell=1$ is needed at 4PN.

  At leading order in $x$ and in the higher-derivative couplings (collectively denoted by $\lambda$ below), the deformed Heun equation reduces to a hypergeometric equation whose general solution is given by
\be\label{ghl}
\varphi_{\inn}(z)\underset{x\,=\,0}{\,=\,}d^{\inn}_1 z^{-\ell} \, _2F_1(-\ell,-\ell;-2 \ell ;z)+d_2^{\inn} z^{\ell+1} \, _2F_1(\ell+1,\ell+1;2 \ell+2;z) +{\cal O}(\lambda)   \, ,
\ee
where $d^\inn_1$ and $d^\inn_2$ are arbitrary constants. For $\ell$ integer, the second solution in (\ref{ghl}) is always singular at $z=1$, the position of the horizon in the uncorrected solution. This means that we must impose $d_2^{\inn}\,=\,0$. The next step is to find how this solution is modified when turning on the higher-derivative couplings $\lambda$. Setting $\ell=1$ and expanding at linear order in $\lambda$, one finds that the solution which is regular at the horizon (suitably normalized) is given by
\be
\varphi_{\rm in}(z)|_{\ell=1}\underset{x\,=\,0}{\,=\,}  \frac{2-z}{2 z} \left(1+\frac{z^3 \lambda _3}{40 M^4}-\frac{11 z^3 \lambda _4}{224
   M^6} +\ldots \right)\, .
 \label{gl1}
\ee
The scalar Love number associated to the $\ell=1$ mode is obtained by expanding the above expression for small $z$ and comparing against the incoming solution in \eqref{eq:incomingsoln_scalar}. Doing so, one obtains
\be
F_1 \underset{x\,=\,0}{\,=\,} \frac{\lambda _3}{40 M^4}-\frac{11 \lambda _4}{224 M^6}\,.
\ee

\subsection{Scalar waveform and energy flux}

Let us compute the scalar waveform and fluxes at 4PN order. To this end, we first notice that we can neglect the higher-derivative corrections to the scalar-wave equation and to the geodesic, which yield contributions at higher PN orders ---namely, 6PN for the cubic terms and 8PN for quartic interactions. Instead, as we just showed, tidal interactions show up at $(2\ell+1)$PN order, thus correcting the waveform already at 3PN order (for the lowest $\ell=1$ mode). The conclusion is that, up to 4PN order, we have that
\begin{equation}
\tilde{Z}_{\ell m}\,=\, \tilde{Z}^{\rm GR}_{\ell m} \left(1+F_\ell  \,z_0^{2\ell+1}+\ldots\right)\,,
\end{equation}
where
\begin{equation}\label{zelm}
\tilde{Z}^{\rm GR}_{\ell m}= \frac{4\pi M \mu \alpha\, C_{\rm in}}{\Gamma}\,\kappa^{\ell m}_0 \,   e^{-\frac{y_0}{2}} \left(1-z_0\right)^{-{x_0\over 2} } z_0^{1+\frac{x_0}{2}}   G_{-}\left(y_0\right) +\ldots
\end{equation}
 is the GR result. We recall that $z_0, x_0$ and $y_0$ are defined in \eqref{eq:PNregime}.
 The first few terms in the PN expansion of the (absolute value of the) scalar waveform are given by
 \begin{equation}\label{Zlmbis}
\begin{aligned}
 &|\tilde Z_{\ell m}^{\rm GR}|  \,= \,
 2\pi \mu \alpha\,\kappa^{\ell m}_0\, {  \Gamma(\ell+1)\over \Gamma(2\ell+2) }  \left(2mv\right)^{\ell+1}   \left[ 1-\left(\frac{3}{2}+\ell+\frac{m^2}{6+4 \ell}\right) v^2 +|m|\pi v^3\right. \\
  &\left. + \left(\frac{3 \ell}{2}-\frac{9}{8}+\frac{(\ell-1)^2 \ell}{2 \ell-1}-\frac{\left(17+7 \ell-2 \ell^2\right) m^2}{12+20 \ell+8
   \ell^2}+\frac{m^4}{120+128 \ell+32 \ell^2}\right) v^4 +\ldots\right]  \, .
\end{aligned}
\end{equation}
Finally, we compute the energy flux, whose expression is \cite{Cipriani:2026myb}
\be\label{eq:defenergysc}
{\diff{\cal E}\over \diff t}   =\sum_{\ell=1}^\infty \sum_{m=-\ell}^\ell {\omega^2 \over \pi } | \tilde{Z}_{\ell m} |^2  \,=\,{\alpha^2 \mu^2 v^{8}\over 3 M^2}  \left(\eta^{\rm GR}+\delta\eta\right) \,.
\ee
Defining $\eta^{\rm GR}\,=\,\sum_{\ell=1}^\infty \sum_{m=1}^\ell \eta_{\ell m}^{\rm GR}$, $\delta \eta\,=\,\sum_{\ell=1}^\infty \sum_{m=1}^\ell \delta\eta_{\ell m}$, we find
\be \label{etalms}
\begin{aligned}
 \eta_{11}^{\rm GR} \,=\,& 1-\frac{26 v^2}{5}+2 \pi  v^3+\frac{1123 v^4}{175} +\ldots\,,  \\[1mm]
  \eta_{22}^{\rm GR}\,=\,&  \frac{16 v^2}{5}-\frac{848 v^4}{35} +\ldots \,,   \hspace{5mm} \eta_{33}^{\rm GR} \,=\,   \frac{2187 v^4}{280} +\ldots  \,, \hspace{5mm} \eta_{31}^{\rm GR} \,=\, \frac{v^4}{1400}  +\ldots   \,,\\[1mm]
\delta\eta_{11}^{\rm GR} \,=\,&  16 \left(\frac{\lambda _3}{40 M^4}-\frac{11 \lambda _4}{224 M^6}\right)v^6\left(1-\frac{13}{5}v^2\right)+\ldots \, ,
 \end{aligned}
 \ee
so that the total energy flux is given by
\begin{equation}\label{dedtscalar}
 \begin{aligned}
   \eta^{\rm GR}\,=\,&1-2 v^2+2 \pi  v^3-10 v^4 +\ldots\,,\\[1mm]
   \delta\eta\,=\,&16 \left(\frac{\lambda _3}{40 M^4}-\frac{11 \lambda _4}{224 M^6}\right)v^6\left(1-\frac{13}{5}v^2\right)+\ldots \, .
 \end{aligned}
 \end{equation}
As anticipated, higher-derivative corrections enter the energy flux at order 3PN, and the correction is proportional to the Love number describing the response of the geometry against  $\ell=1$ perturbations. Our final formula for the energy flux \eqref{dedtscalar} correctly reduces to (5.14) in \cite{Bini:2016egn} once the higher-derivative couplings are set to zero.

\section{Gravitational waveform and energy/angular momentum fluxes}\label{sec:waveform+flux}

In this section we compute the gravitational waveform  as well as the energy and angular momentum fluxes at 6PN order. As in the previous section, we will consider a point-like source on circular motion around the black hole background \eqref{eq:bkgmetric}.
This section is organized as follows. First, in sections~\ref{sec:incomingsolodd} and \ref{sec:incomingsoleven} we compute two independent homogeneous solutions of the odd and even master equations. A linear combination of these two gives the incoming one, which is needed to compute the waveform. As it turns out, the ratio between the coefficients of the linear combination is related to the tidal Love numbers of the geometry, whose computation is revisited in section~\ref{sec:PNexpansion}. In section \ref{sec:waveform_circ} we compute the stress-energy tensor of a point-like source on circular motion, and use it to evaluate the source terms in the master equations. Finally, in sections~\ref{sec:waveform} and \ref{sec:energy} we compute the waveform and fluxes, respectively.

Before delving into the details, we find convenient to denote by $\Psi_{\ell m}\,=\, \{\Psi_{\odd, \ell m}, \Psi_{\ev, \ell m}\}$ and $S_{\ell m}\,=\, \{S_{\odd, \ell m}, S_{\ev, \ell m}\}$ the even/odd master variables and source terms that enter in the modified Regge-Wheeler \eqref{eq:mastereqodd} and Zerilli \eqref{eq:mastereven} equations.
The solutions to those equations satisfying incoming boundary conditions at the horizon and outgoing at infinity are given by
\begin{equation}\label{eq:inhomogeneous_soln}
\Psi_{\ell m}(r) \,=\,\int_{r_+}^{\infty}G\left(r, r'\right)S_{\ell m}(r')\,\diff r'\,,
\end{equation}
where
\be\label{eq:G_odd}
G(r,r')\,=\,
{1\over W}   \left\{
\begin{array}{ccc}
  \Psi^{\rm in}( r')\,\Psi^{\rm out} ( r)  & ~~~~~~~&  r'< r \\
 \Psi^{\rm in} (r)\,\Psi^{\rm out} ( r')  & ~~~~~~~ &   r< r' \\
\end{array}
\right.\,,
\ee
and
\begin{equation}
W\,=\, f N \left(\Psi^{\inn}\frac{\diff \Psi^\out}{\diff r}-\Psi^{\out}\frac{\diff \Psi^\inn}{\diff r}\right)\,.
\end{equation}
$\Psi^\inn$ and $\Psi^\out$ are solutions of the homogeneous equation satisfying incoming boundary conditions at the horizon and outgoing at infinity. A normalization is chosen so that
\begin{equation}\label{eq:bdrycondinf0}
\begin{aligned}
     \Psi^{\rm in}(r) \,\underset{r\to \infty}{\longrightarrow}\,&  B^{\rm in}_{+}  \,   e^{{\rm i} \omega r_*} +  {1\over 2 {\rm i} \omega} \,  e^{-{\rm i} \omega r_*} \,,
     \qquad\qquad
 \Psi^{\rm out}(r) \underset{r\to \infty}{\longrightarrow}    \,   e^{{\rm i} \omega r_*}  \,,
\end{aligned}
\end{equation}
where we recall that $r_*$ is the tortoise coordinate \eqref{eq:tortoise}. This implies that
\begin{equation}
W\,=\,1\,.
\end{equation}
The waveform $Z_{\ell m}$ is again defined as the amplitude of the solution \eqref{eq:inhomogeneous_soln} in the asympotic region, where it reduces to
\begin{equation}
\Psi_{\ell m} (r)\underset{r\to \infty}{\longrightarrow}Z_{\ell m} \, e^{\ii\omega r_*}\,,
\end{equation}
with
\begin{equation}\label{eq:waveform_general}
Z_{\ell m} \,=\,\int_{r_+}^\infty  \Psi^\inn (r') \, S_{\ell m}(r')\, \diff r'\,.
\end{equation}
  In the next sections, we compute the two ingredients entering this formula: the incoming solutions and the source terms.

 \subsection{Odd incoming solution}
 \label{sec:incomingsolodd}

Consider the homogenous odd master equation,
\be\label{eq:mastereqodd0}
\frac{\diff}{\diff r}\left[f N \frac{\diff \Psi_{\rm odd}}{\diff r}\right] + \frac{1}{fN}\left(\omega^2 - f V_{\rm odd} \right)\Psi_{\rm odd}\,=\, 0\,,
\ee
where the effective potential $V_{\rm odd}$ was given in (\ref{vodd}), (\ref{vodd3}) and (\ref{vodd4}). Following the same steps as in the previous section, we make use of the dimensionless variables in \eqref{eq:definitions} and define a function $G_{\odd}(y)$ as follows
\begin{equation}\label{eq:Godd}
\Psi_\odd\left(\frac{2M y}{x}\right)\,=\,  e^{-\frac{y}{2}}\left(1-\frac{x}{y}\right)^{-\frac{x}{2}}y^{-\tfrac{x}{2}}\,G_{\odd}(y) \, .
\end{equation}
Inserting \eqref{eq:Godd} into \eqref{eq:mastereqodd0},  one finds a second-order differential equation for $G_{\odd}(y)$, which takes the form of a modified CHE:
\begin{equation}\label{eq:modifiedCHE}
\left({\cal L}_{\rm CHE}+ {\cal L}_{\lambda, \odd} \right)G_{\odd}(y)\,=\, 0\,.
\end{equation}
The operator ${\cal L}_{\rm CHE}$ is again the one given in \eqref{lCHE}, but it is important to bear in mind that the dictionary between the CHE parameters, $m_1, m_2, m_3, u$, and those appearing in the master equations are modified with respect to the scalar case studied in the previous section. Concretely, now we have that
\be\label{modd}
  m_1=\frac{4+x}{2}  ~ , \qquad  m_2=\frac{-4+x}{2}~ , \qquad  m_3=-\frac{x}{2} ~ , \qquad  u \,=\, \left(\ell+\frac{1}{2}\right)^2 + \frac{x}{2}+ \frac{x^2}{4} \,.
\ee
In turn, ${\cal L}_{\lambda, \odd}$ denotes the contributions linear in the higher-derivative couplings, which we do not write explicitly as they are lengthy uninformative expressions. All that matters for us is that in the strict $x\to 0$ limit the equation becomes hypergeometric, allowing us to apply the strategy of \cite{Cipriani:2025ikx} in order to construct the solutions as a series expansion in $x$. The technical steps are the same as in the scalar case. Namely, we write an ansatz of the form
\begin{equation}
  G_{\odd, \alpha} \left( y\right) =  P_{\odd}\left( y\right) \, H_{\odd, \alpha}\left( y\right) {+} \widehat{P}_{\odd}\left(y\right) \,  y \frac{\diff H_{\odd, \alpha}(y)}{\diff y}\,, \hspace{1cm} \alpha\,=\, \{+, -\}\,,
\end{equation}
where
\begin{equation}
H_{\odd, \alpha}(y)\,=\,y^{\frac{1}{2}-\alpha a_\odd+\tfrac{x}{2}} \,
   _1F_1\left(\ft{1}{2}-\alpha a_\odd+\tfrac{x}{2};1-2\alpha
   a_\odd;y\right)\,,
\end{equation}
and
\begin{equation}
\begin{aligned}
P_{\odd}(y) \,=\,& 1+\sum_{i=1}^k   \sum_{j=0}^{i-1}   c^{\odd}_{ij} \, x^i y^{j-i} \,,  \hspace{1cm} \widehat{P}_{\odd}(y)\,=\, \sum_{i=1}^k   \sum_{j=0}^{i-1}   \widehat{c}^{\odd}_{ij} \, x^i y^{j-i} \,. \end{aligned}
 \end{equation}
The parameter $a_\odd$ is related to $u$ (and thus $\ell$) by
\begin{equation}\label{eq:ui}
u \,=\, \left(\ell+\frac{1}{2}\right)^2+ \frac{x}{2}+\frac{x^2}{4}\,=\, a^2_\odd+\sum_{i=1}^k u^{\odd}_i x^i\,,
\end{equation}
where $u_i^{\odd}$ are constant coefficients (depending on $a_\odd$) that one must find together with  $c^{\odd}_{ij}$ and ${\widehat c}^{\odd}_{ij}$ by solving \eqref{eq:mastereqodd0} order by order in $x$. As it turns out, for the 6PN calculation that we have in mind, it suffices to find the solution up to order $x^6$. In order to display the explicit expressions for the polynomials, it is convenient to write them as
\begin{equation}
 P_{\odd}(y)\,=\, P_{\rm {GR}}(y) +\delta P_{\odd}(y)\, , \hspace{1cm}  {\widehat{P}}_{\odd}(y)\,=\, {\widehat{P}}_{\rm {GR}}(y) +\delta {\widehat{P}}_{\odd}(y)\, ,
\end{equation}
with $\delta P_{\odd}$ and $\delta {\widehat{P}}_{\odd}$ accounting for the contributions from the higher-derivative corrections. The complete solutions up to order $x^6$ are provided in the ancillary Mathematica file. In order to avoid reporting lengthy expressions, here we follow the presentation in the scalar case and organize the terms in a PN expansion \eqref{eq:PNregime}, providing explicitly the GR contribution up to 2.5PN,
\begin{equation}\label{eq:polynomialsGR}
\begin{aligned}
P_{\rm GR}\,=\,& 1+\frac{x}{2 y}+\frac{\left(313-104 a_{\rm GR}^2+16 a_{\rm GR}^4\right) x^2}{128 \left(a_{\rm GR}^2-1\right) y^2}-\frac{\left(315+248 a_{\rm GR}^2+112 a_{\rm GR}^4\right) x^2}{128 \left(a_{\rm GR}^2-1\right) \left(4a_{\rm GR}^2-1\right) y}+\ldots\,, \\[1mm]
\widehat{P}_{\rm GR}  \,=\,& \frac{\left(4 a_{\rm GR}^2-17\right) x  }{2\left(1-4 a_{\rm GR}^2\right) y}+\frac{\left(523+232 a_{\rm GR}^2-80 a_{\rm GR}^4\right) x^2  }{64 \left(1-5 a_{\rm GR}^2+4 a_{\rm GR}^4\right)
   y^2} +\frac{\left(17-4 a_{\rm GR}^2\right)^2 x^2  }{8 \left(1-4 a_{\rm GR}^2\right)^2 y}+\ldots\,.
\end{aligned}
\end{equation}
Next, we display the corrections up to 6PN order, which is all we will need for the aim of this paper:
\begin{equation}
\begin{aligned}\label{eq:polynomialsD}
\delta P_\odd\,=\,&\frac{\lambda _3}{M^4}  \left( -\frac{21 x^6 \left(12 a_{\rm GR}^2-20 \ell (\ell+1)+517\right)}{512 \left(a_{\rm GR}^2-9\right)y^6} +\frac{9 x^5}{2 \left(-9+4
   a_{\rm GR}^2\right)  y^3} \right)+\ldots\,,\\[1mm]
\delta {\widehat P}_{\odd}\,=\,& \frac{\lambda _3}{M^4}  \left[ \frac{3 x^6 \left(12 a_{\rm GR}^2+28 \ell(\ell+1)  -983\right)}{256 \left(a_{\rm GR}^2-9\right) y^6}+\frac{9 x^5}{2 \left(-9+4a_{\rm GR}^2\right)  y^3} \right] +\ldots\,.
\end{aligned}
\end{equation}
Finally, we can invert \eqref{eq:ui} to find $a_\odd$ as a function of $\ell$. Again, the result can be expressed as
\begin{equation}
a_{\odd}\,=\, a_{\rm GR} + \,\delta a_{\odd}\,,
\end{equation}
where
\begin{equation}\label{eq:aGR}
a_{\rm GR}\,=\, \ell+\frac{1}{2}+\frac{\left(24+13 \ell+28 \ell^2+30
   \ell^3+15 \ell^4\right) x^2}{8 \ell (1+2 \ell)
   \left(-3+\ell+8 \ell^2+4 \ell^3\right)}+{\cal O}(x^4)\,,
\end{equation}
and
\begin{equation}
\delta a_\odd\,=\,-\frac{9\lambda_3 ( 353 \ell^6+1059 \ell^5-2102 \ell^4-5969
   \ell^3-2031 \ell^2+1130 \ell+18900 ) x^6}{128M^4 (-1+\ell^2) \ell (2+\ell) (4 \ell^2-25) (-9+4 \ell^2) (-1+4 \ell^2)
   (7+2 \ell) }+{\cal O}(x^8)\, .
\end{equation}
As we can see, only the cubic terms give a contribution at the order we are working. The quartic corrections enter at 8PN, and so we can consistently neglect them.

The incoming solution is a linear combination of the two independent solutions we have found,
\begin{equation}\label{eq:oddincoming}
\Psi_\odd^\inn\left(\frac{2M y}{x}\right)\,=\,C_\odd^{\inn}\, e^{-\frac{y}{2}}\left(1-\frac{x}{y}\right)^{-\frac{x}{2}}y^{-\tfrac{x}{2}}\,\left[G_{\odd, -}(y)+x^{2a_\odd}\, F_{\odd, \, \ell}\,G_{\odd, +}(y)\right]\, ,
\end{equation}
where $C^{\inn}_{\odd}$ is an overall normalization which we fixed when imposing \eqref{eq:bdrycondinf0}. Using the asymptotic behavior of the hypergeometric functions given in \eqref{eq:conform}, one finds
\be
\Psi_{\rm odd}(r) \underset{r\to \infty}{\longrightarrow}  \,C_\odd^{\inn}\, e^{-{\rm i} \omega r_*}  x^{-\tfrac{x}{2}}\,\left(B^{\odd}_{-- }+x^{2a_\odd}\, F_{\odd, \, \ell}\,B^{\odd}_{+- }\right) +B^+_{\rm in}  e^{{\rm i} \omega r_*} \,,
\ee
so that
\be\label{eq:Coddin}
 C_{\rm odd}^{\rm in} = {2M x^{ {x\over 2}-1 } \over B^{\rm odd}_{--}  \left(  1  + x^{2a_{\rm odd}} F_{\odd, \,\ell} \ft{B^{\rm odd}_{+-}}{B^{\rm odd}_{--} }\right)}\,,
\ee
where
\be
B_{\alpha\beta}^{\odd}\,=\,e^{\frac{\ii \pi\left(1-\beta\right)}{2}\left(\frac{1+x}{2}-\alpha a_\odd\right)} {\Gamma(1-2 \alpha a_\odd)\over \Gamma\left( \ft12-\alpha a_\odd+\beta\ft{x}{2} \right) }\,, \hspace{1cm} \alpha, \beta \,=\,\{+, -\}\, .
\ee
In turn, $F_{\odd, \, \ell}$ is a constant to be fixed by imposing incoming boundary conditions at the horizon. Splitting the GR and higher-derivative contributions as
\begin{equation}
F_{\odd, \, \ell} \,=\, F_{{\rm GR}, \, \ell} + \delta F_{\odd, \, \ell}\,,
\end{equation}
one has that \cite{Cipriani:2025ikx}
\begin{equation}\label{eq:F_GR}
F_{{\rm GR}, \, \ell}\,=\,- \frac{F_{-+}}{F_{++}}\, {\rm exp}\left(   \sum_{i=1}^\infty   \partial_{a_{\rm GR}} u^{\rm GR}_i {x^i\over i} \right)\,,
\end{equation}
where
\begin{align}
F_{\alpha\beta}  =  \frac{\Gamma(1{+}2\alpha a_{\rm GR}) \Gamma (- \beta x) }{\Gamma \left(\ft{1}{2}{+}\alpha a_{\rm GR}{-}\beta \left(2+\ft{x}{2}\right)  \right)
	\Gamma \left(\ft{1}{2}{+}\alpha a_{\rm GR}{-}\beta  \left(-2+\ft{x}{2}\right)  \right)}\,, \hspace{1cm} \alpha, \beta \,=\,\{+, -\}\, .
\end{align}
 As we further explain in section~\ref{sec:PNexpansion}, the quantity $F_{\rm GR, \ell}$ corresponds to the dynamical Love numbers of the Schwarzschild black hole. As it is well known, their static limit vanishes,
 $\lim_{x\to 0}F_{\rm GR, \ell}\,=\,0$. This will no longer be true when including the higher-derivative corrections, whose contribution $\delta F_{\odd, \ell}$ is computed in section~\ref{sec:PNexpansion}. Leaving $\delta F_{\odd, \ell}$ unspecified for the time being, and splitting the GR plus higher-derivative contributions as
\begin{equation}
\Psi^{\inn}_{\odd}\,=\,\Psi^{\inn,  \rm GR}_{\odd} + \delta\Psi^{\inn}_{\odd}\,,
\end{equation}
we find
\begin{equation}
\begin{aligned}
\Psi_{\odd}^{\inn,  \rm GR} \,=\,& C^{\inn}_{\odd}\, y^{\ell+1}\left[1+\frac{\left(\frac{2}{\ell}-\frac{\ell}{2}\right) x}{y}+\frac{y^2}{24+16 \ell}-\frac{\left(\ell^2+\ell-4 \right) x}{4 \ell
   (1+\ell)}+\frac{ (\ell-3) (\ell+1) \left(\ell^2-4\right)  x^2}{4 \ell (2 \ell-1) y^2}\right.\\[1mm]
   &\left.+\frac{\left(12+14 \ell+5 \ell^2-\ell^3\right) x
   y}{48 \ell+80 \ell^2+32 \ell^3}+\frac{y^4}{128 \left(15+16 \ell+4 \ell^2\right)}+\dots\right]\, ,
\end{aligned}
\end{equation}
and
\begin{equation}\label{eq:dPsiodd}
\begin{aligned}
\delta\Psi^{\inn}_\odd    \,=\,& C^{\inn}_{\odd}
\,y^{\ell+1}\left\{ \frac{\lambda_3}{M^4}\left[\frac{3 \left(-100-4 \ell+5 \ell^2\right) x^6}{16(-5+2 \ell) y^6}+\frac{9 x^5}{8 (-1+\ell) y^3}+\dots\right]\right.\\[1mm]
 &\left.+\delta F_{\odd, \,\ell}\left(\frac{x}{y}\right)^{2\ell+1} \left[1+\frac{\left(-3+2 \ell+\ell^2\right) x}{2 (1+\ell) y}+\frac{y^2}{8(1-2 \ell)}+\dots \right]\right\}\,.
 \end{aligned}
\end{equation}
 Let us remark that the contributions in the second line modify the GR solution at $(2\ell+1)$PN order. Since we work at 6PN, in what follows we will ignore these corrections for all the modes except for $\ell=2$, for which they eventually give the leading 5PN correction.

\subsection{Even incoming solution}
\label{sec:incomingsoleven}

We consider next the homogenous even master equation,
\be\label{eq:mastereven0}
 \frac{\diff }{\diff r}\left[N f \frac{\diff\Psi_{\ev}}{\diff r}\right] +\frac{1}{N f}\left(\omega^2 - f V_{\ev}\right)\Psi_{\ev}\,=\, 0\,,
\ee
with $V_{\ev}$ given by (\ref{veven}), (\ref{veven3}), (\ref{veven4}).  This equation is not a CHE, even in GR. The way to circumvent this issue in GR is to make use of the so-called Chandrasekhar transformation \cite{Chandrasekhar:1985kt}, which maps a solution to the  Zerilli equation to
a solution of the Regge-Wheeler equation preserving the boundary conditions. However, it is known that higher-derivative corrections break isospectrality, so no relation between the solutions of the even and odd master equations is expected beyond GR. Still, Chandrasekhar’s trick is useful in order to come up with an ansatz to solve the corrected even master equation. In order to describe the ansatz, we introduce an auxiliary function $\Psi_{\rm aux}(r)$, related to $\Psi_\ev(r)$ by
\begin{equation}\label{chandra}
\Psi_{\ev}(r)\,=\, \left[\frac{\nu  (\nu +2)}{6}+\frac{12 M^2 (r-2 M)}{r^2 (6 M+\nu  r)} \right] \,\Psi_{\rm aux}(r) +  2Mf(r)  \,\frac{\diff\Psi_{\rm aux}(r)}{\diff r}\,.
\end{equation}
Next, we introduce a function $G_{\ev}(y)$ as follows,
\begin{equation}\label{eq:Gev}
\Psi_{\rm aux}\left( \frac{2My}{x}\right)\,=\, e^{-\frac{y}{2}}\left(1-\frac{x}{y}\right)^{-\frac{x}{2}}y^{-\tfrac{x}{2}}\,G_{\ev}(y) \, .
\end{equation}
Following the same steps as in the scalar and odd cases, we make an ansatz for $G_{\ev}(y)$ based on a hypergeometric function:
\begin{equation}\label{Geven}
  G_{\ev, \alpha} \left( y\right) =  P_{\ev}\left( y\right) \, H_{\ev, \alpha}\left( y\right) {+} \widehat{P}_{\ev}\left(y\right) \,  y \frac{\diff H_{\ev,\alpha}\left( y\right)}{\diff y}\,, \hspace{1cm} \alpha\,=\, \{+, -\}\,,
\end{equation}
where
\begin{equation}
H_{\ev, \alpha}(y)\,=\,y^{\frac{1}{2}-\alpha a_\ev+\tfrac{x}{2}} \,
   _1F_1\left(\ft{1}{2}-\alpha a_\ev+\tfrac{x}{2};1-2\alpha
   a_\ev;y\right)\,,
\end{equation}
\begin{equation}
\begin{aligned}
P_{\ev}(y) \,=\,& 1+\sum_{i=1}^k   \sum_{j=0}^{i-1}   c^{\ev}_{ij} \, x^i y^{j-i} \,,  \hspace{1cm} \widehat{P}_{\ev}(y)\,=\, \sum_{i=1}^k   \sum_{j=0}^{i-1}   \widehat{c}^{\ev}_{ij} \, x^i y^{j-i} \,,
\end{aligned}
\end{equation}
 and
\begin{equation}
 \left(\ell+\frac{1}{2}\right)^2+\frac{x}{2}+\frac{x^2}{4}\,=\, a^2_\ev+\sum_{i=1}^k u_i^\ev x^i\,.
\end{equation}
Substituting this ansatz in the homogeneous master equation \eqref{eq:mastereven0} and expanding perturbatively in $x$ and in the higher-derivative couplings, one obtains again a set of algebraic equations for the coefficients $c_{ij}^\ev, {\widehat c}_{ij}^\ev, u^\ev_i$. Solving them order by order, one gets the two independent solutions to the even master equation \eqref{eq:mastereven0}. Expressing the solution as GR plus higher-derivative corrections,
\begin{equation}
 P_{\ev}\,=\, P_{\rm GR}+\delta P_{\ev}\,, \hspace{1cm}   {\widehat P}_{\ev}\,=\, {\widehat P}_{\rm GR}+\delta {\widehat P}_{\ev}\,,
\end{equation}
we find that $P_{\rm GR}$ and ${\widehat P}_{\rm GR}$ are the same as in the odd case \eqref{eq:polynomialsGR}, as expected. In turn, the corrections in the even sector differ from the odd ones \eqref{eq:polynomialsD}, and are given by
\begin{equation}
\begin{aligned}
\delta P_\ev\,=\,&\frac{\lambda_3}{M^4}\left[\frac{9 x^5}{2 y^5}-\frac{9 x^5}{8 y^4}+\frac{9 x^5}{\left(18-8 a_{\rm GR}^2\right) y^3}\right.\\[1mm]
&\left.-\frac{3 \left(1713663-413148 a_{\rm GR}^2+43376 a_{\rm GR}^4-18752 a_{\rm GR}^6+1536 a_{\rm GR}^8\right) x^6}{256 \left(-9+a_{\rm GR}^2\right) \left(9-40 a_{\rm GR}^2+16
   a_{\rm GR}^4\right) y^6}+\dots\right]\,,\\[1mm]
\delta {\widehat P}_\ev\,=\,&\frac{\lambda_3}{M^4}\left[\frac{9 x^5}{4 y^5}+\frac{9 x^5}{\left(18-8 a_{\rm GR}^2\right) y^3}\right.\\[1mm]
&\left.-\frac{3 \left(205875+44724 a_{\rm GR}^2-32816 a_{\rm GR}^4+2240 a_{\rm GR}^6\right) x^6}{128 \left(-9+a_{\rm GR}^2\right) \left(-9+4 a_{\rm GR}^2\right)
   \left(-1+4 a_{\rm GR}^2\right) y^6}+\dots\right]\, .
\end{aligned}
\end{equation}
Finally, we provide the expression for $a_\ev$ as a function of $\ell$, which reads
\begin{equation}
 a_\ev\,=\,a_{\rm GR}+\delta a_\ev\,,
\end{equation}
where $a_{\rm GR}$ is given in \eqref{eq:aGR} and
\begin{equation}
\begin{aligned}
\delta a_{\ev}\,=\,& \frac{9 \left({-}18900{+}9230 \ell{+}5811 \ell^2{-}6491 \ell^3{-}2378 \ell^4{+}1041 \ell^5{+}347 \ell^6\right)
   x^6 \eta ^8  \lambda_3 }{128 \ell \left({-}35{+}4 \ell{+}4 \ell^2\right)
   \left(30{-}23 \ell{-}19 \ell^2{+}8 \ell^3{+}4 \ell^4\right) \left({-}3{-}5 \ell{+}10 \ell^2{+}20 \ell^3{+}8
   \ell^4\right) M^4} \\
   &+{\cal O}(x^8)\,.
\end{aligned}
\end{equation}

The even incoming solution is found by applying the Chandrasekhar-like transformation \eqref{chandra} to $\Psi^{\inn}_{\rm aux}(r)$, where
\begin{equation}\label{eq:evenincoming}
\Psi_{\rm aux}^{\inn}\left(\frac{2My}{x}\right)\,=\, C_{\ev}^{\inn} \,e^{-\frac{y}{2}}\left(1-\frac{x}{y}\right)^{-\frac{x}{2}}y^{-\tfrac{x}{2}}\,\left[G_{\ev, -}(y)+F_{\ev, \, \ell}\,x^{2a_\ev}\,G_{\ev, +}(y)\right]  \,.
\end{equation}
The coefficient $F_{\ev,\ell}$ is fixed by demanding incoming boundary conditions, and it is given by
\begin{equation}
 F_{\ev, \, \ell}\,=\, F_{{\rm GR}, \, \ell}+\delta F_{\ev, \, \ell}\,,
\end{equation}
where $F_{{\rm GR}, \, \ell}$ is given in \eqref{eq:F_GR} and $\delta F_{\ev, \, \ell}$ are the corrections to the Love numbers, which we determine in the next section.  The normalization  $C^{\inn}_{\ev}$ is determined as in the odd case, using the asymptotic behaviour of the wave function,
\be
\Psi_{\ev }(r) \underset{r\to \infty}{\longrightarrow}  \,  C^\inn_\ev  x^{-\tfrac{x}{2}} \left[\frac{(\ell (\ell^2{-}1)(\ell{+}2)}{6}{-}\frac{x}{2}\right] \, e^{-{\rm i} \omega r_*}  \,\left(B^\ev_{-- }{+}x^{2a_\ev}\, F_{\ev, \, \ell}\,B^\ev_{+- }\right) {+}B^+_{\rm in}  e^{{\rm i} \omega r_*} \,,
\ee
 and imposing  (\ref{eq:bdrycondinf0}). This yields
 \be\label{eq:Cinev}
 C_{\ev}^{\rm in} = { 2M x^{ {x\over 2}-1 } \over B^{\ev}_{--}  \left(1  + x^{2a_{\ev}} F_{\ev, \, \ell} \ft{B^{\ev}_{+-}}{B^{\ev}_{--} }\right) \left[\frac{(\ell (\ell^2-1)(\ell+2)}{6}-\frac{x}{2}\right]}\,,
  \ee
where
\be
B_{\alpha\beta}^{\ev}\,=\,e^{\frac{\ii \pi\left(1-\beta\right)}{2}\left(\frac{1+x}{2}-\alpha a_\ev\right)} {\Gamma(1-2 \alpha a_\ev)\over \Gamma\left( \ft12-\alpha a_\ev+\beta\ft{x}{2} \right) }\,, \hspace{1cm} \alpha, \beta \,=\,\{+, -\}\, .
\ee
Let us note that the additional factor in the denominator of \eqref{eq:Cinev} with respect to \eqref{eq:Coddin} comes from the Chandrasekhar-like transformation.

Separating the GR and higher-derivative contributions as before,
\begin{equation}
\Psi^{\inn}_\ev\,=\, \Psi^{\inn, \rm GR}_\ev + \delta \Psi^{\inn}_\ev\,,
\end{equation}
we find that
\begin{equation}
\begin{aligned}
\Psi^{\inn, \rm GR}_\ev  \,=\,&   C^\inn_\ev \frac{ \ell (\ell^2-1)(\ell+2)}{6} y^{\ell+1}\left[1+\frac{y^2}{24+16 \ell}-\frac{\left(-4-4 \ell-6 \ell^2+\ell^3+\ell^4\right) x}{2 (-1+\ell) \ell (2+\ell) y}\right.\\[1mm]
&{-}\frac{\left(-4+\ell+\ell^2\right) x}{4 \ell (1+\ell)}+\frac{\left(12-4 \ell+16 \ell^2+21 \ell^3+4 \ell^4-\ell^5\right) x y}{16 (-1+\ell) \ell (1+\ell) (2+\ell) (3+2 \ell)}+\frac{y^4}{128 \left(15+16
   \ell+4 \ell^2\right)}\\[1mm]
&\left.+\frac{\left(24-16 \ell+24 \ell^2-16 \ell^3+37 \ell^4-4 \ell^5-14 \ell^6+\ell^8\right) x^2}{4 \ell (-1+2 \ell) \left(-2+\ell+\ell^2\right)^2 y^2}+\dots\right]\,,
\end{aligned}
\end{equation}
and
\begin{equation}
\begin{aligned}\label{dpsievin}
\delta \Psi^{\inn}_\ev\,=\,& C^{\inn}_\ev \frac{ \ell (\ell^2-1)(\ell+2)}{6} y^{\ell+1}\left\{\frac{\lambda_3}{M^4}\left[\frac{9 (3+\ell) x^5}{4 y^5}+\frac{9 \left(-17-4 \ell+\ell^2\right) x^5}{32 \left(-3+\ell+2 \ell^2\right) y^3}\right.\right.\\[1mm]
&\left.+\frac{3 \left(240+272 \ell+360 \ell^2+54 \ell^3-41 \ell^4-12 \ell^5\right) x^6}{16 \ell (2+\ell) (-5+2 \ell) y^6}+\dots\right]\\[1mm]
&\left.+\delta F_{\ev, \, \ell} \left(\frac{x}{y}\right)^{2\ell+1}\left[1+\frac{y^2}{8(1-2 \ell)}+\frac{\left(6+7 \ell+3 \ell^2-3 \ell^3-\ell^4\right) x}{\left(4+2 \ell-4 \ell^2-2 \ell^3\right) y}+\dots\right]\right\}\, .
\end{aligned}
\end{equation}

\subsection{Love numbers}\label{sec:PNexpansion}

In this section we compute the coefficients $F_\ell =\{F_{\odd, \ell}, F_{\ev, \ell}\}$. The calculation we have to do was already explained in some detail for scalar perturbations in section~\ref{sec:scalarLove}, so let us be brief here. These coefficients compute the dynamical Love numbers of the black hole \eqref{eq:bkgmetric}. Indeed, we have that the incoming solutions in the \emph{near-zone} $x\ll z \ll 1$ can be approximated by
\begin{equation}\label{eq:incomingsoln_evodd2}
\Psi^{\rm in}(z)  \,\underset{x\ll z \ll 1}{=}    z^{-\hat{\ell}-1} (1+\ldots )\left[1+F_{\ell} \,z^{2\hat{\ell}+1} (1+\ldots )  \right] \,,
\end{equation}
where $\hat{\ell}\,=\,\ell + {\cal O}\left(M^2\omega^2\right)$ and the dots represent polynomial corrections in $z$. For our purposes here, we need only the zero-frequency limit ($x\to 0$) of these quantities for the $\ell=2$ mode. These have been computed in the previous literature, \cite{Cardoso:2018ptl, Cano:2025zyk}, but let us revisit this calculation anyway in order to make this section more self-contained.  All we have to do is to find the PN expansion ($x\ll 1$, $z$ finite) of the incoming solution. We consider the odd and even cases separately.

\subsection*{Odd case}

Ignoring higher-derivative corrections, one finds that the most general solution to \eqref{eq:mastereqodd0} in the $x \to 0$ limit is
\begin{equation}
\Psi^{\inn}_{\odd}\left(z\right) \underset{x\,=\,0}{\,=\,} \frac{d^{\inn}_1}{z^3}   +d^{\inn}_2\,\frac{3 z^4+4 z^3+6 z^2+12 z+12 \log (1-z)}{z^3} +{\cal O}(\lambda)\,,
\end{equation}
where $d_1^{\inn}$ and $d^{\inn}_2$ are arbitrary coefficients. Obviously, the second solution is not regular at the horizon ---placed at $z=1$ in the uncorrected geometry---, and so we must set $d_2^{\inn}\,=\,0$. Next we consider the higher-derivative corrections. Denoting the correction to the solution as
\be
\Psi^{\rm in}_{\rm odd} \underset{x\,=\,0}{\,=\,} {1\over z^3} +   \delta \Psi^{\rm in}_{\rm odd}(z)\,,
\ee
one finds it obeys the following differential equation,
\begin{equation}
\begin{aligned}
 &\left(1-z\right) z^4 \frac{\diff^2\delta \Psi^{\rm in}_{\rm odd}}{\diff z^2} + \left(2-3 z\right) z^3 \frac{\diff\delta \Psi^{\rm in}_{\rm odd}}{\diff z}+ 3 \left(-2+z\right) z^2\delta\Psi^{\rm in}_{\rm odd} \\[1mm]
& \quad + \frac{\lambda _3}{ M^4} 3 z^5 \left(-33+41 z\right) -\frac{\lambda _4}{4M^6} 9 z^7 \left(1104-2756 z+1677 z^2\right)+\frac{108\tilde{\lambda}_4}{M^6} z^7\,=\, 0\, .
\end{aligned}
\end{equation}
Suitably choosing the normalization constant $d_1^\inn$, the solution which is regular at the horizon can be written as
\begin{equation}\label{eq:PsioddPN}
\begin{aligned}
\Psi^{\rm in}_{\rm odd}(z)  \underset{x\,=\,0}{\,=\,}& {1\over z^3} \left[  1+\frac{\lambda _3}{M^4} z^5 \left(-\frac{15 }{2}+\frac{41 z}{4}\right) +\frac{ \tilde{\lambda}_4}{M^6} z^5\left(\frac{36}{5} + 6 z + \frac{36 z^2}{7}\right) \right. \\[1mm]
   & \left. +\frac{\lambda
   _4}{M^6} z^5 \left(-\frac{162}{25} -\frac{27 z}{5}-\frac{162 z^2}{35} + \frac{1989 z^3}{20}-\frac{1677 z^4}{20}\right) \right]\, .
\end{aligned}
\end{equation}
We can now compare this solution with the one obtained in section~\ref{sec:incomingsolodd} ---concretely, with \eqref{eq:dPsiodd}--- in the regime in which both PN and PM expansions are valid, the near-zone $x\ll z \ll 1$. Matching the coefficients of the $z^5$ terms, we obtain that $\delta F_{\odd, 2}$ is given by:\footnote{Actually, we can also compare the coefficients of $z^6$ in \eqref{eq:PsioddPN} and \eqref{eq:dPsiodd}, and we see that they perfectly match once we have imposed \eqref{Fodd}.}
\be\label{Fodd}
\delta F_{\odd, 2}\underset{x\,=\,0}{\,=\,}  -\frac{15}{2}\frac{\lambda_3}{M^4}-\frac{162}{25}\frac{\lambda_4}{M^6}+\frac{36}{5}\frac{{\tilde\lambda}_4}{M^6} \, .
\ee
This result agrees with the ones obtained in previous literature  \cite{Cardoso:2018ptl,Cano:2025zyk}. In order to facilitate the comparison, we remind the reader that the Love numbers $k_\ell^\pm$ in those references are for solutions of the Teukolsky equation, related to those  in the RWZ formalism via the rescaling \cite{Cano:2025zyk}
\be
\delta F_{\odd, \, \ell}\underset{x\,=\,0}{\,=\,} 2^{-2\ell} \frac{(\ell+1)(\ell+2)}{\ell(\ell-1)} k_\ell^-\, .
\ee
Specifying to $\ell=2$ gives $ \delta F_{\odd, \, 2}\underset{x\,=\,0}{\,=\,}\frac{3}{8}k_2^-$ in agreement with \cite{Cano:2025zyk}, see Table~1.

\subsection*{Even case}

Proceeding as before, one finds that the regular solution in the even case is given by
\begin{equation}
\begin{aligned}
{\Psi}^{\inn}_{\ev}(z)\underset{x\,=\,0}{\,=\,}&  \frac{4+6 z-3 z^3}{z^3\left(4+3 z\right)}+\frac{\lambda_3}{M^4}\frac{z^2 \left(2784-224 z-3258 z^2-657 z^3+666 z^4+108 z^5\right) }{8 (4+3 z)^2} \\[1mm]
&-\frac{\lambda_4}{M^6}\frac{z^5}{100 (4+3 z)^2} \left(-24192-44352 z-44496 z^2+37674 z^3+38428 z^4\right.\\[1mm]
&\left.-43350 z^5-27900 z^6+12300 z^7+2025 z^8\right) \, .
\end{aligned}
\end{equation}
Expanding for small $z$, one finds
\begin{equation}
\begin{aligned}
{\Psi}^\inn_\ev(z)\underset{x\,=\,0}{\,=\,}&\frac{1}{z^3}+\frac{3}{4z^2}-\frac{9 }{16z}-\frac{21}{64}+\frac{63 z}{256}+\left(-\frac{189}{1024}+\frac{87}{4}\frac{\lambda_3}{M^4}+\frac{378}{25}\frac{\lambda_4}{M^6}\right)
   z^2\\[1mm]
&+\left(\frac{567}{4096}-\frac{275}{8} \frac{\lambda_3}{M^4}+\frac{126}{25}\frac{\lambda_4}{M^6}\right) z^3+{\mathcal O}\left(z^4\right)\,.
\end{aligned}
\end{equation}
Matching this with the incoming solution \eqref{dpsievin} in the near-zone we find that
\begin{equation}
\delta F_{\ev, 2} \underset{x\,=\,0}{\,=\,}21\frac{\lambda_3}{M^4}+\frac{378}{25}\frac{\lambda_4}{M^6}\,,
\end{equation}
in agreement with \cite{Cardoso:2018ptl, Cano:2025zyk} after the identification $\delta F_{\ev, 2} \underset{x\,=\,0}{\,=\,}\frac{3}{8}k_2^+$.

\subsection{Energy momentum tensor for a test particle on circular motion}
\label{sec:waveform_circ}

The energy-momentum tensor for a massive point-like particle is
 \be
T^{\mu\nu}=\mu \int {d\tau  \over   \sqrt{-g}} \frac{\diff\bar{x}^\mu}{\diff\tau} \frac{\diff\bar{x}^\nu}{\diff\tau}\delta^{(4)}(x-\bar{x}(\tau))\,.
\ee
If the particle moves along a circular geodesic of radius $r_0$ along the equator, we have that
\be
 {\diff\bar x^\mu\over \diff \tau} =\left(  \Gamma , 0, 0, \Gamma \, \Omega_{\rm orb}  \right)
\,, \hspace{1cm}
  \Gamma \,=\,    {E\over \mu f(r_0)N(r_0)^2} \,, \hspace{1cm} \Omega_{\rm orb}= {J f(r_0)N(r_0)^2  \over E r_0^2} \,,
\end{equation}
where $E$ and $J$ were given in \eqref{eq:circular_orbits}. Hence,
\be\label{eq:Tmunucirc1}
T^{\mu\nu}\,=\,{\cal T}\, \frac{\diff {\bar x}^\mu}{\diff t} \frac{\diff {\bar x}^\nu}{\diff t} \,, \hspace{1cm} {\cal T}\,=\, { \mu \Gamma\over N r^2 }    \delta(r-r_0) \delta(\cos\theta) \delta(\phi-\Omega_{\rm orb} t)\,,
\ee
and the non-trivial components of $T_{\mu\nu}$ are \be\label{eq:Tmunucirc2}
T^{00}\,=\,{\cal T}   \,, \hspace{1cm}  T^{\phi 0} \,=\,\Omega_{\rm orb} {\cal T} \,, \hspace{1cm}  T^{\phi \phi} \,=\, \Omega_{\rm orb}^2  {\cal T}\,.
\ee
The different components $\mathcal{T}_{ab}$, $\mathcal{T}_{a}$, $\mathcal{P}$, $\mathcal{Q}$, $\mathcal{S}_{a}$, $\mathcal{S}_{2}$ entering in the source terms can be obtained by Fourier-transforming $T_{\mu\nu}$ and projecting it on the different spherical harmonics. Using the relevant orthogonality relations, we get
\begin{equation}
\begin{aligned}
\mathcal{T}_{ab}&=\frac{1}{2\pi}\int dt e^{i \omega t}\int d\Omega\, \bar{Y}^{\ell m} T_{ab}\, ,\\
\mathcal{T}_{a}&=\frac{1}{2\pi (\nu+2)}\int dt e^{i \omega t}\int d\Omega\, \bar{Y}^{A\,\ell m} T_{aA}\, ,\\
\mathcal{P}&=\frac{1}{4\pi}\int dt e^{i \omega t}\int d\Omega\, \bar{Y}^{\ell m} \Omega^{AB}T_{AB}\, ,\\
\mathcal{Q}&=\frac{1}{\pi\nu(\nu+2)}\int dt e^{i \omega t}\int d\Omega\, \bar{Y}^{AB\, \ell m}T_{AB}\, ,\\
\mathcal{S}_{a}&=\frac{1}{2\pi(\nu+2)}\int dt e^{i \omega t}\int d\Omega\, \bar{X}^{A\,\ell m} T_{aA}\, ,\\
\mathcal{S}_{2}&=\frac{1}{\pi\nu(\nu+2)}\int dt e^{i \omega t}\int d\Omega\, \bar{X}^{AB\, \ell m}T_{AB}\, .
\end{aligned}
\end{equation}

\paragraph{$T_{\mu\nu}$ components evaluated for circular orbits.} Using \eqref{eq:Tmunucirc1} and \eqref{eq:Tmunucirc2}, one finds that the non-vanishing components of the stress-energy momentum for a particle on circular motion are given by
\begin{equation}\label{eq:Scirc}
\begin{aligned}
{\cal T}_{00} \,=\,&  \frac{E f(r_0) N(r_0)}{r_0^2} \,  \kappa^{\ell m}_0 \delta(r-r_0) \delta(\omega -m \Omega_{\rm orb})\,, \\[1mm]
 {\cal T}_{0} \,=\,& \frac{\ii m\, J f(r_0) N(r_0) }{(\nu+2) r^2_0}\,\kappa^{\ell m}_0 \delta(r-r_0) \delta(\omega -m \Omega_{\rm orb})\,,\\
{\cal P} \,=\,& \frac{J^2 f(r_0) N(r_0)}{2 E r^2_0}\, \kappa^{\ell m}_0\left(\ft{\pi}{2}, 0\right)  \delta(\omega -m \Omega_{\rm orb})\,,\\[1mm]
\cal Q \,=\,& \frac{J^2 f(r_0) N(r_0)  \left(\nu+2-2 m^2\right)}{E r^2 \nu(\nu+2)}\, \kappa^{\ell m}_0 \delta(r-r_0) \delta(\omega -m \Omega_{\rm orb})\,,\\[1mm]
{\cal S}_0 \,=\,& -  \frac{J f\left(r_0\right)N\left(r_0\right)}{(\nu+2)r^2_0  }  \kappa^{\ell m}_1    \delta(r-r_0) \delta(\omega -m \Omega_{\rm orb})\,, \\[1mm]
{\cal S}_2\,=\,&  - \frac{ 2 {\rm i} m  J^2 f(r_0) N(r_0)  }{E r^2_0  \nu(\nu+2)}   \kappa^{\ell m}_1     \delta(r-r_0)\delta(\omega -m \Omega_{\rm orb})\,,
\end{aligned}
\end{equation}
with
\be
\kappa^{\ell m}_0={\bar Y}^{\ell m}\left(\ft{\pi}{2}, 0\right) \,, \hspace{1cm} \kappa^{\ell m}_1= \partial_\theta {\bar Y}^{\ell m}\left(\theta, 0\right)\big|_{\theta =\ft{\pi}{2}}\, .
\ee
We recall again that the expressions for $E$ and $J$ are given in \eqref{eq:circular_orbits}. Next, we use \eqref{eq:Scirc} to evaluate the source terms in \eqref{eq:mastereqodd} and \eqref{eq:mastereven}, which we do not write explicitly. Plugging the resulting expressions for $S(r)$ in the general formula for the waveform \eqref{eq:waveform_general}, integrating by parts and using the fact that $\Psi_{\inn}(r)$ is a solution of the homogeneous equation, we can express the waveform as
\begin{equation}\label{eq:waveform_circ}
Z_{\rm \ell m} \,\equiv\, \delta(\omega-m\Omega_{\rm orb})\tilde{Z}_{\ell m}\,,
\end{equation}
with
\be\label{tildeZdef}
\tilde{Z}_{\ell m}  \,=\, A_0  \Psi^\inn\left(r_0\right)+ A_1 \, r_0\,  \frac{\diff\Psi^{\inn}(r_0)}{\diff r}\,.
\ee
  Here $A_0$, $A_1$ depends on $\ell$, $m$ and the velocity.  Their complete expressions  in the odd/even cases can be found in Appendix \ref{app:coefficients}.  Up to 6PN order, one finds
\beaq
 && A_{0, \odd}\,=\,\,\frac{4\mu v^3 \left(1-2 v^2\right) \kappa_1^{\ell m}  }{M\nu  (2+\nu ) \sqrt{1-3 v^2} }\left(1+  \frac{2\lambda_3}{M^4}\left(355+72 m^2\right) v^{12}+ \ldots   \right)\,,\nn\\
 &&  A_{1, \odd}\,=\,\,\frac{4\mu v^3 \left(1-2 v^2\right)  \kappa_1^{\ell m}  }{M\nu  (2+\nu ) \sqrt{1-3 v^2} }\left(1+ \frac{134\lambda_3}{M^4}v^{12}+ \ldots \right)\,,\\
 &&  A_{0, \ev}  \,=\,\,{-}\frac{2\mu v^2  \nu  \kappa_0^{\ell m}  }{M \left(6 v^2{+}\nu \right)^2  \sqrt{1{-}3 v^2}}\left[1{-}\frac{\left(2 m^2{+}\nu{-}4 \right)v^2 }{2{+}\nu}{-}\frac{24 \left(m^2{-}2\right) v^4}{\nu  (2{+}\nu)}
  {-}\frac{12
   \left(6 m^2{-}6{+}\nu \right)v^6}{\nu ^2 (2{+}\nu )}\right] \nn\\
   &&\cdot \left\{1{+}   \frac{\lambda_3}{M^4}\left[ {-}\frac{288(\nu{-}3)v^{10}}{2{+}\nu}{+} \frac{24 v^{12} \left(37 \nu ^3-2 \left(12 m^2+73\right) \nu ^2{+}8 \left(9 m^2{-}109\right) \nu
   {-}144\right)}{\nu  (\nu {+}2)^2 } {+} \ldots \right]  \right\}\,, \nn\\[1mm]
&& A_{1, \ev} \,=\,-\frac{4 \mu v^2  (1-2v^2)^2  \kappa_0^{\ell m}  }{M(2{+}\nu ) \left(6 v^2+\nu \right)  \sqrt{1{-}3 v^2} }\left\{1+ \frac{\lambda_3}{M^4}\left[ -36 (\nu -2) v^{10} -\frac{24 (5 \nu +18) v^{12}}{\nu } +\ldots \right]  \right\}\,.\nn
\eeaq

\subsection{Waveform}
\label{sec:waveform}

At this stage, we have computed all the ingredients we need to obtain the waveform \eqref{eq:waveform_circ}. We are interested in its absolute value, which we write as
\begin{equation}
|\tilde{Z}_{\ell m}|\,=\, |\tilde{Z}_{\ell m}|_N ( \Gamma_{{\rm GR}, \ell m}+  \delta\Gamma_{\ell m} ) \,,
\label{gammaGR}
\end{equation}
where $|\tilde{Z}_{\ell m}|_N$ is the GR result at leading order in the PN expansion,
\begin{align}\label{eq:leadingZlmeven}
\left|\tilde{Z}^\ev_{\ell m}\right|_{N}&\,=\,\frac{2\mu\, \Gamma(\ell-1)}{\Gamma(2\ell+2)} \,|\kappa_0^{\ell m}| \,\left(2|m|v\right)^\ell \, ,\\[1mm]
\left|\tilde{Z}^\odd_{\ell m}\right|_{N}&\,=\, \frac{4\mu\,  \Gamma(\ell-1)}{(\ell+1)\Gamma(2\ell+2)} \,| \kappa_1^{\ell m}|\, (2|m|)^\ell\, v^{\ell+1}\, ,
\label{eq:leadingZlmodd}
\end{align}
and $\Gamma_{{\rm GR}, \ell m}$ and $\delta\Gamma_{\ell m}$ are the PN corrections coming from GR and from the higher-derivative terms, respectively. We will present them momentarily. First, let us notice that $Y^{\ell m}\left(\tfrac{\pi}{2},0\right)$ vanishes for $\ell+m$ odd, while its first derivative vanishes for $\ell+m$ even, which implies
\be\label{eq:Zevenodd}
\tilde{Z}_{\ell m}=\left\{
\begin{array}{lll}
 \tilde{Z}^{\ev}_{\ell m}   & ~~~~~~~&  \ell+m \in 2\mathbb{Z} \\
 \tilde{Z}^{\odd}_{\ell m}   & ~~~~~~~ &   \ell+m\in 2\mathbb{Z}+1 \\
\end{array}
\right.\,.
\ee
As in the rest of the section, we provide the GR result, $\Gamma_{{\rm GR}, \ell m}$, up to order 2PN with respect to the leading term. In turn, the higher-derivative contribution $\delta \Gamma_{\ell m}$ will be given up to 6PN, so we will provide the leading and next-to-leading orders. This will allow us to see the interplay between the different contributions to the waveform from the higher-derivative interactions: those associated to the boundary conditions versus those coming from the explicit corrections to the wave equations and geodesic motion. At the GR level, one finds
\begin{equation}\label{GammaGR}
\begin{aligned}
\Gamma^{\ev}_{\rm GR}\,=\,&
1+\left(\frac{1}{2}-\ell-\frac{(9+\ell) m^2}{6+10 \ell+4 \ell^2}\right) v^2+|m| \pi v^3+v^4 \left(\frac{15-41 \ell+66 \ell^2-32 \ell^3+8
   \ell^4}{8-24 \ell+16 \ell^2} \right. \\[1mm]
   & \left. {+}\frac{\left(6 \ell{-}32{-}51 \ell^2{-}23 \ell^3+3 \ell^4+\ell^5\right) m^2}{4 (\ell-1) \ell (1+\ell)^2
   (2+\ell)}+\frac{\left(50+19 \ell+\ell^2\right) m^4}{8 (1+\ell) (2+\ell) (3+2 \ell) (5+2 \ell)}\right)+\ldots\,, \\[1mm]
\Gamma^{\odd}_{\rm GR}\,=\,& 1+\left(\frac{1}{2}+\frac{2}{\ell}-\ell-\frac{(4+\ell) m^2}{12+14 \ell+4 \ell^2}\right) v^2+ |m| \pi v^3+v^4\left(-\frac{8+33 \ell+2 \ell^2-24
   \ell^3+8 \ell^4}{8 \ell-16 \ell^2} \right. \\[1mm]
& \left. -\frac{\left(72+104 \ell+53 \ell^2+3 \ell^3-2 \ell^4\right) m^2}{24 \ell+52 \ell^2+36 \ell^3+8
   \ell^4}+\frac{(6+\ell) m^4}{8 \left(30+47 \ell+24 \ell^2+4 \ell^3\right)}\right) +\ldots\,.
   \end{aligned}
\end{equation}
In turn, we can express the contributions from the higher-derivative terms as follows
\begin{equation}\label{eq:deltaGamma}
\begin{aligned}
\delta\Gamma^{\ev}_{\ell m}\,=\,& 2k_2^+  v^{10}\left(1+\frac{3}{2}v^2+\dots\right)\delta_{\ell 2}\delta_{m2}+\frac{2 \ell (43+8 \ell) v^{12}}{-5+2 \ell}\frac{\lambda_3}{M^4}+\dots\,, \\[1mm]
\delta\Gamma^{\odd}_{\ell m}\,=\,&-3k_2^- v^{10}\left(1+\frac{8v^2}{3}+\dots\right)\delta_{\ell 2}\delta_{m1}-\frac{10 (1+\ell) (-29+8 \ell) v^{12}}{-5+2 \ell}\frac{\lambda_3}{M^4}+\dots\,.
\end{aligned}
\end{equation}
where we recall that
\begin{equation}\label{eq:Love}
k_2^+\,=\,28\frac{\lambda_3}{M^4}+\frac{1008}{25}\frac{\lambda_4}{M^6}\,,\hspace{1cm}
k_2^- \,=\,-20\frac{\lambda_3}{M^4}-\frac{432}{25}\frac{\lambda_4}{M^6}+\frac{96}{5}\frac{{\tilde\lambda}_4}{M^6}\,.
\end{equation}
This emphasizes our main conclusion, which is that the leading corrections to the waveform coming from higher-derivative terms arise at 5PN, and are proportional to the static Love numbers for the $\ell=2$ modes.

\subsection{Energy and angular momentum fluxes}
\label{sec:energy}

 The energy and angular momentum fluxes at infinity can be computed out of the waveform by using the results of \cite{Martel:2005ir}. Taking into consideration \eqref{eq:Zevenodd}, one has\footnote{The extra factor of $\left(8\pi\right)^2$ with respect to formula (6.16) in  \cite{Martel:2005ir} is due to the normalization of the perturbation, $h_{\mu\nu}^{\rm MP}\,=\,8\pi h_{\mu\nu}^{\rm here}$. Recall that we fixed $G=1$.}
\begin{equation}\label{energyflux}
{\diff{\cal E}\over \diff t} \,=\, {\left(8\pi\right)^2\over 64 \pi}    \sum_{\ell \,=\, 2}^\infty \sum_{m\,=\,-\ell}^\ell \omega^2 \ell (\ell^2-1)(\ell+2) |\tilde Z_{\ell m} |^2  \,=\,\left( {\diff{\cal E}\over \diff t}\right)_N  \left( \eta^{\rm GR}+\delta\eta\right) \,,
\end{equation}
where
\begin{equation}
\left( {\diff {\cal E}\over \diff t}\right)_N\,=\,{32 \mu^2 v^{10}\over 5M^2}\,
\end{equation}
is the leading term in GR, coming from the $\ell=2$ even modes, and\footnote{In this section we give the contribution to the energy flux from GR at order 2PN. In appendix~\ref{app:etaGR}, we give $\eta^{\rm GR}$ up to $v^{12}$ to connect the result of general relativity with $\delta\eta$. Our results, obtained with the formalism of \cite{Cunningham:1978zfa}, are the same of those obtained with the formalism of Teukolsky, so the interested reader can see Appendix E of \cite{Mino:1997bx} for the contribution of each single mode to the total energy emission.}
 \beaq
  \eta^{\rm GR} &\,=\,&  1-\frac{1247 v^2}{336}+4
   \pi  v^3-\frac{44711 v^4}{9072}+\ldots \,,
   \label{energyPN}
   \\[1mm]
  \delta\eta  &\,=\,&
  4\left(28\frac{\lambda_3}{M^4}+\frac{1008}{25}\frac{\lambda_4}{M^6} \right) v^{10}   - \left( 726\frac{\lambda_3 }{M^4}    +\frac{4152}{25}\frac{\lambda_4}{M^6} +\frac{16}{5}\frac{\tilde\lambda_4}{M^6} \right)v^{12}{+}\ldots\,.
  \label{deltaeta}
\eeaq
Making use of the the static tidal Love numbers \eqref{eq:Love}, we can write $\delta \eta$ as
\begin{equation}
\delta\eta\,=\, 4 k_2^+ v^{10}-v^{12} \left(\frac{k_2^-}{6}+ \frac{88k_2^+}{21} +\frac{612\lambda_3}{M^4}\right)\, .
\end{equation}
Finally, we conclude computing the flux of angular momentum, which is defined as \cite{Martel:2005ir}
\begin{equation}\label{Jflux}
{\diff{J}\over \diff t} \,=\, {\left(8\pi\right)^2\over 64 \pi}    \sum_{\ell \,=\, 2}^\infty \sum_{m\,=\,-\ell}^\ell m\omega \ell (\ell^2-1)(\ell+2) |\tilde Z_{\ell m} |^2\,=\,\frac{1}{\Omega_{\rm orb}} {\diff{\cal E}\over \diff t}  \,.
\end{equation}
Then, we simply have that
\begin{equation}\label{Jflux2}
\frac{\diff J}{\diff t} \,=\, \frac{32\mu^2v^7}{5M}   \left(\eta^{\rm GR}+\delta \eta^J\right)\,, \hspace{1cm} \delta \eta^J  \,=\,  \delta \eta +70 v^{12}{\lambda_3\over M^4}+\ldots\, .
 \end{equation}
 We have thus obtained that the leading higher-derivative corrections to the energy and the angular momentum fluxes appear up at 5PN order, and are proportional to the Love number $k^+_2$ given in \eqref{eq:Love}. Moreover, we have verified that our results correctly reproduce those obtained using the Teukolsky formalism \cite{Mino:1997bx} when ignoring the contribution from the higher-derivative interactions. In contrast, our predictions for the contributions from the higher-derivative interactions disagrees with the previous literature, which finds these corrections appear at high orders in the PN expansion \cite{Endlich:2017tqa,Cardoso:2018ptl,Sennett:2019bpc, Brandhuber:2024bnz,Brandhuber:2024lgl,Liu:2024atc}. We further comment on this in our conclusions.


\section{Numerical results}
\label{sec:numericalresults}
In order to verify the analytic results and extend them to higher values of $v$, we have performed a numerical calculation of the amplitudes $\tilde{Z}_{\ell m}$ defined in \eqref{tildeZdef}.
We recall that \eqref{tildeZdef} has been obtained by assuming the normalization \eqref{eq:bdrycondinf0} on the  $\Psi^{\rm in}$ and $\Psi^{\rm out}$ solutions, which fixes the Wronskian to $W=1$. For the numerical computation, we consider instead the normalization
\begin{align}
\Psi^{\rm in}=e^{-i\omega r_{*}}\quad  \text{when}\quad r_{*}\to -\infty\, ,\\
\Psi^{\rm out}=e^{+i\omega r_{*}}\quad  \text{when}\quad r_{*}\to +\infty\, ,
\end{align}
\textit{i.e.}, the normalization of the incoming solution is not fixed at infinity but at the horizon. As a result, $W\neq 1$, and we simply have to add a global $1/W$ factor to the right hand side of \eqref{tildeZdef}.
In order to obtain the $\Psi^{\rm in}$ and $\Psi^{\rm out}$ solutions and the Wronskian, we first consider the expansions of these functions near $r=r_{+}$ and $r=\infty$, which take the form
\begin{align}
\Psi^{\rm in}=\left(\frac{r}{r_{+}}-1\right)^{-\frac{i\omega}{4\pi T}}\left[1+\sum_{n=1}^{\infty}a_n\left(\frac{r}{r_{+}}-1\right)^n\right]\quad  \text{when}\quad r \to r_{+}\, ,\\
\Psi^{\rm out}=e^{i\omega r }\left(\frac{r}{2M}\right)^{2iM\omega}\left[1+\sum_{n=1}^{\infty}b_n\left(\frac{2M}{r}\right)^n\right]\quad  \text{when}\quad r\to +\infty\, \, ,
\end{align}
where $T=f'(r_{+})N(r_+)/(4\pi)$ is the Hawking temperature.
The coefficients $a_n$ and $b_n$ can be found analytically by solving the equation order by order in either $(r-r_{+})$ or in $1/r$, respectively. The idea is then to use these expansions to set the boundary conditions for the numerical integration at some finite $r$.
For $\Psi^{\rm in}$, we use the expansion with $n=3$ to determine the boundary conditions for the numerical method at $r/r_{+}-1=10^{-4}$. For $\Psi^{\rm out}$ we include $n=5$ terms and set the boundary conditions at $r=100 \ell^2/|\omega|$ (a radius much longer than the wavelength and than the size of the black hole). For the numerical routine we use the \texttt{NDSolve} function provided in Mathematica, and we work with extra precision in order to be able to resolve even tiny changes in the waveform. In particular, we set \texttt{AccuracyGoal}$=$\texttt{PrecisionGoal}$=$24 and \texttt{WorkinPrecision}$=$39. We solve the problem for GR and for small values of the higher-derivative couplings so that the linear approximation in the coupling is accurate.
For the results we present next, we employed $\lambda_{3}/M^4=10^{-4}$, $\lambda_{4}/M^6=10^{-4}$, $\tilde{\lambda}_{4}/M^6=10^{-4}$ and extract the linear coefficient of the corrections to $\tilde{Z}_{\ell m}$ by subtracting the GR value. We have verified that, for these values of the couplings, the effects of the higher-derivative corrections are indeed very approximately linear. We now discuss our results.

\begin{figure}[h!]
\center
	\includegraphics[width=0.6\textwidth]{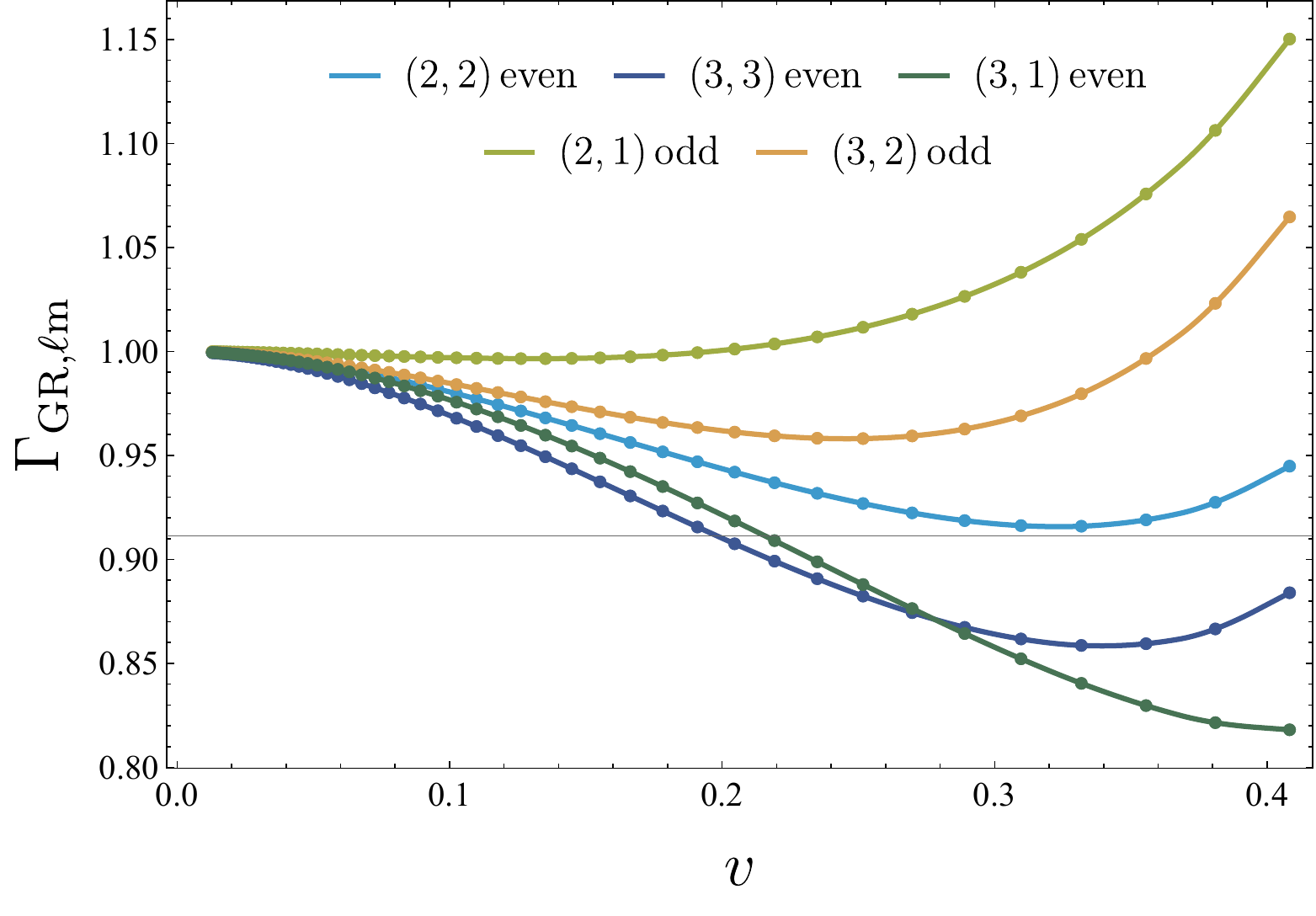}
	\caption{Amplitude of each $\ell m$ mode in the GW signal for GR, normalized by the leading PN behavior \eqref{eq:leadingZlmeven}, \eqref{eq:leadingZlmodd}. Each dot is a numerically computed value, and the lines are an interpolating function.}
	\label{figure:ZlmGR}
\end{figure}

\subsection*{General Relativity}
As a check of our computations, we show first the results for GR.  In Fig.~\ref{figure:ZlmGR} we show the relative amplitudes $\Gamma_{\rm GR, \ell m}=\left|{ Z}_{\ell m}\right|/\left|{Z}_{\ell m}\right|_{N}$ (introduced in \eqref{gammaGR}) for a few values of $\ell,m$. As we can see, all these quantities approach $1$ for small velocities and exhibit a mild variation as $v$ is increased. By doing a fit for small values of $v$, we can extract numerically the subleading coefficients in the PN expansion of $\Gamma_{\rm GR, \ell m}$. Although this process does not yield fully accurate values for the PN coefficients --- because a fit is not equivalent to a series expansion --- the results that we obtain are consistent with the analytic formulae in \eqref{eq:deltaGamma}. In particular, we have checked that the numerically obtained coefficients for the first few PN corrections approach the analytic values as we reduce the range of the fit to smaller $v$.

\begin{figure}[t!]
\center
	\includegraphics[width=0.6\textwidth]{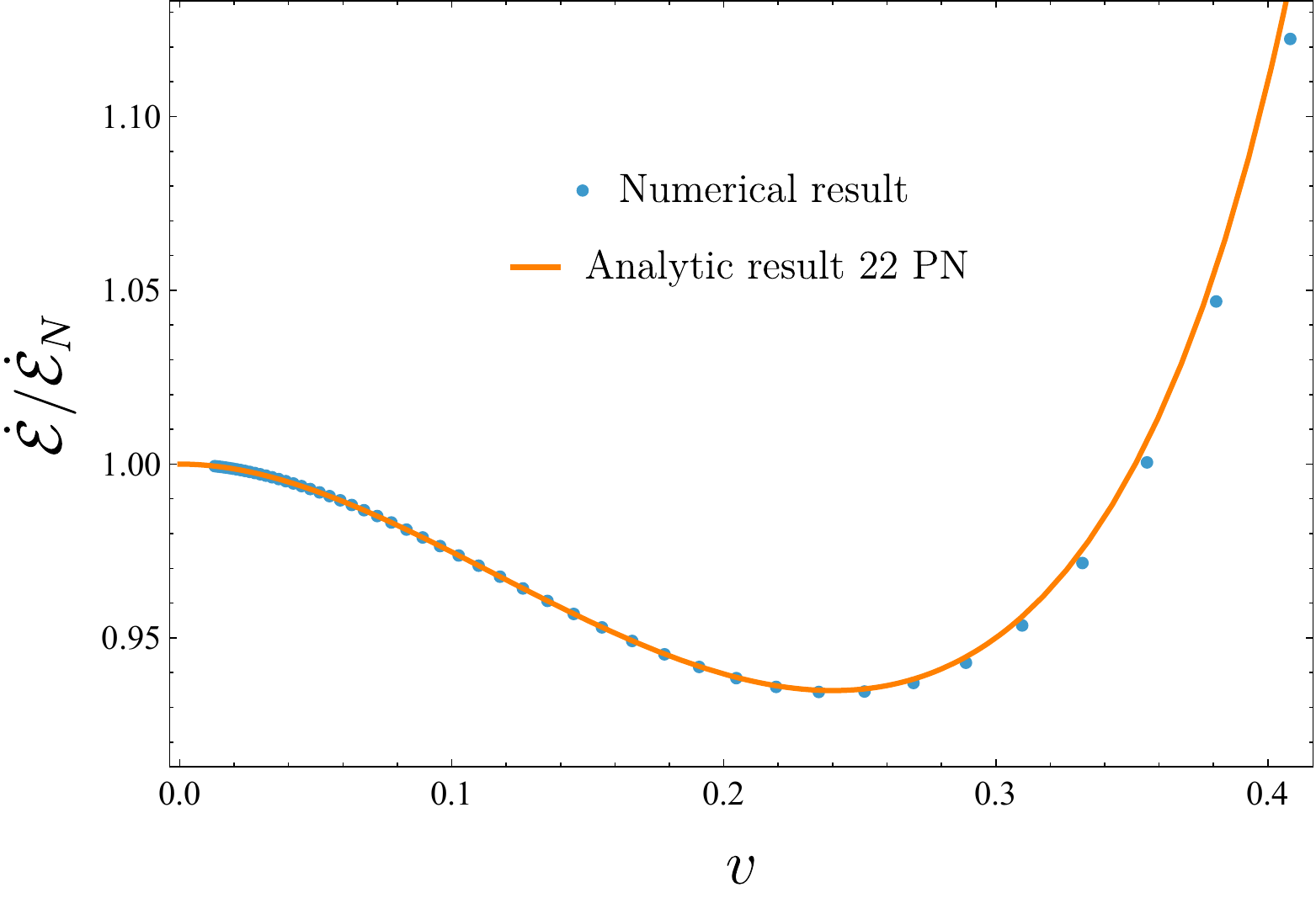}
	\caption{Energy flux in GR, normalized by the leading-order quadrupole term $\dot{\E}_{N}$. We show the numerical results obtained from summing all the modes with $\ell\le 4$ along with analytical results to order 22PN from \cite{Fujita:2012cm}. }
	\label{figure:powerGR}
\end{figure}

An explicit comparison with analytic results is given in Fig.~\ref{figure:powerGR}, where we show the power emitted in GR relative to the leading order $\left(\frac{d\E}{dt}\right)_{N}$. To compute the power numerically, we sum the modes with $\ell\le 4$. We compare the result with the analytic formula for $\frac{d\E}{dt}$ at 22PN \cite{Fujita:2012cm,Cipriani:2025ikx}, finding excellent agreement for $v\lesssim 0.4$. The discrepancy at higher $v$ is likely due to the truncation of the expansion and to the fact that we are only including up to $\ell=4$ modes in the numerical computation.

\subsection*{Higher-derivative corrections}
We now focus on the higher-derivative corrections to the amplitudes, denoted by $\delta \Gamma_{\ell m}$ in \eqref{gammaGR}. At linear order in the coupling constants, we split the different contributions as
\begin{equation}
\delta \Gamma_{\ell m}=\frac{\lambda_{3}}{M^4}\delta \Gamma_{3,\ell m}+\frac{\lambda_{4}}{M^6}\delta \Gamma_{4,\ell m}+\frac{\tilde{\lambda}_{4}}{M^6}\delta \tilde{\Gamma}_{4,\ell m}\, ,
\end{equation}
for both the odd and even sectors, so that each $\delta \Gamma_{i,\ell m}$ coefficient is dimensionless.

\begin{figure}[t!]
\center
	\includegraphics[width=0.48\textwidth]{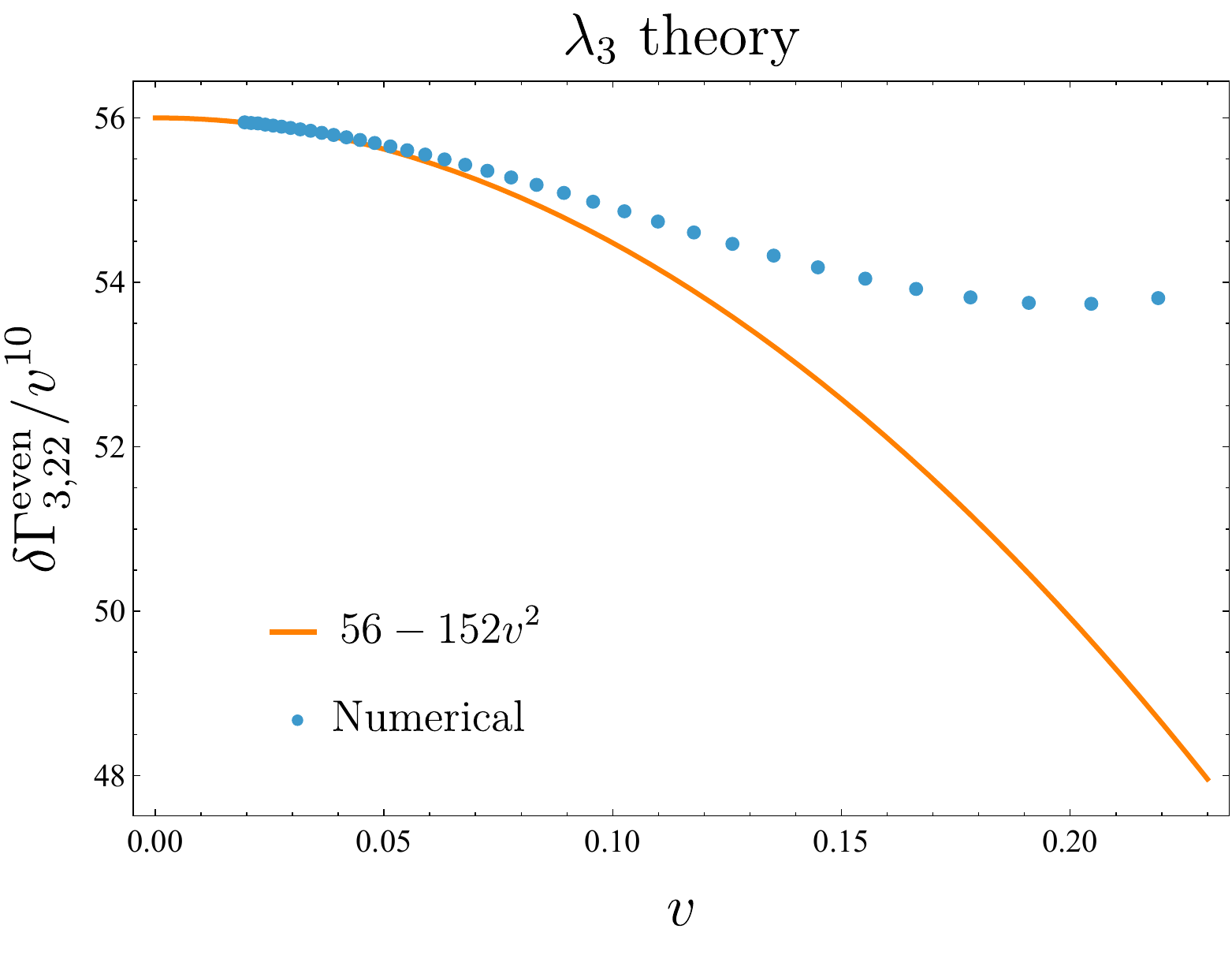}
	\includegraphics[width=0.48\textwidth]{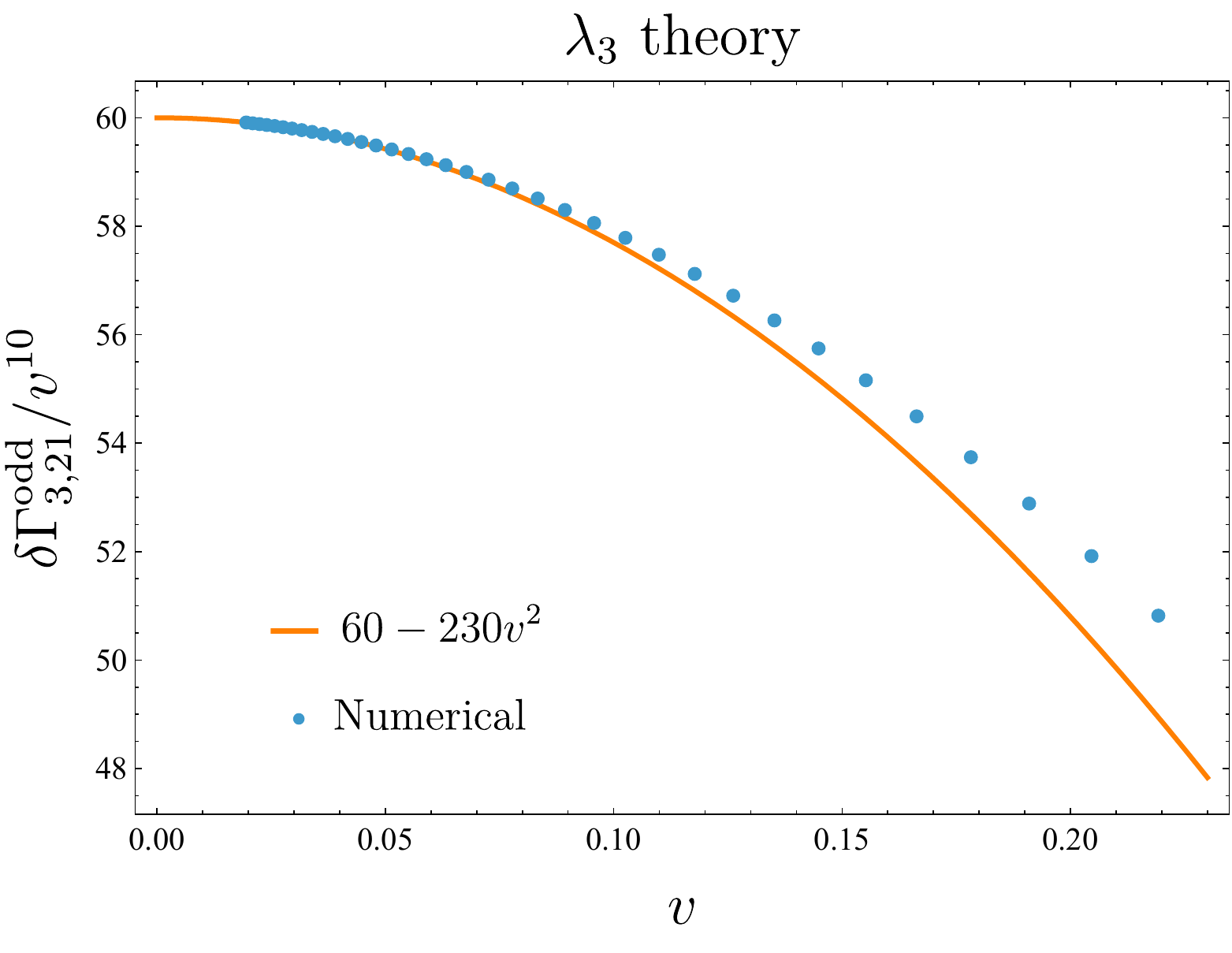}
    \includegraphics[width=0.48\textwidth]{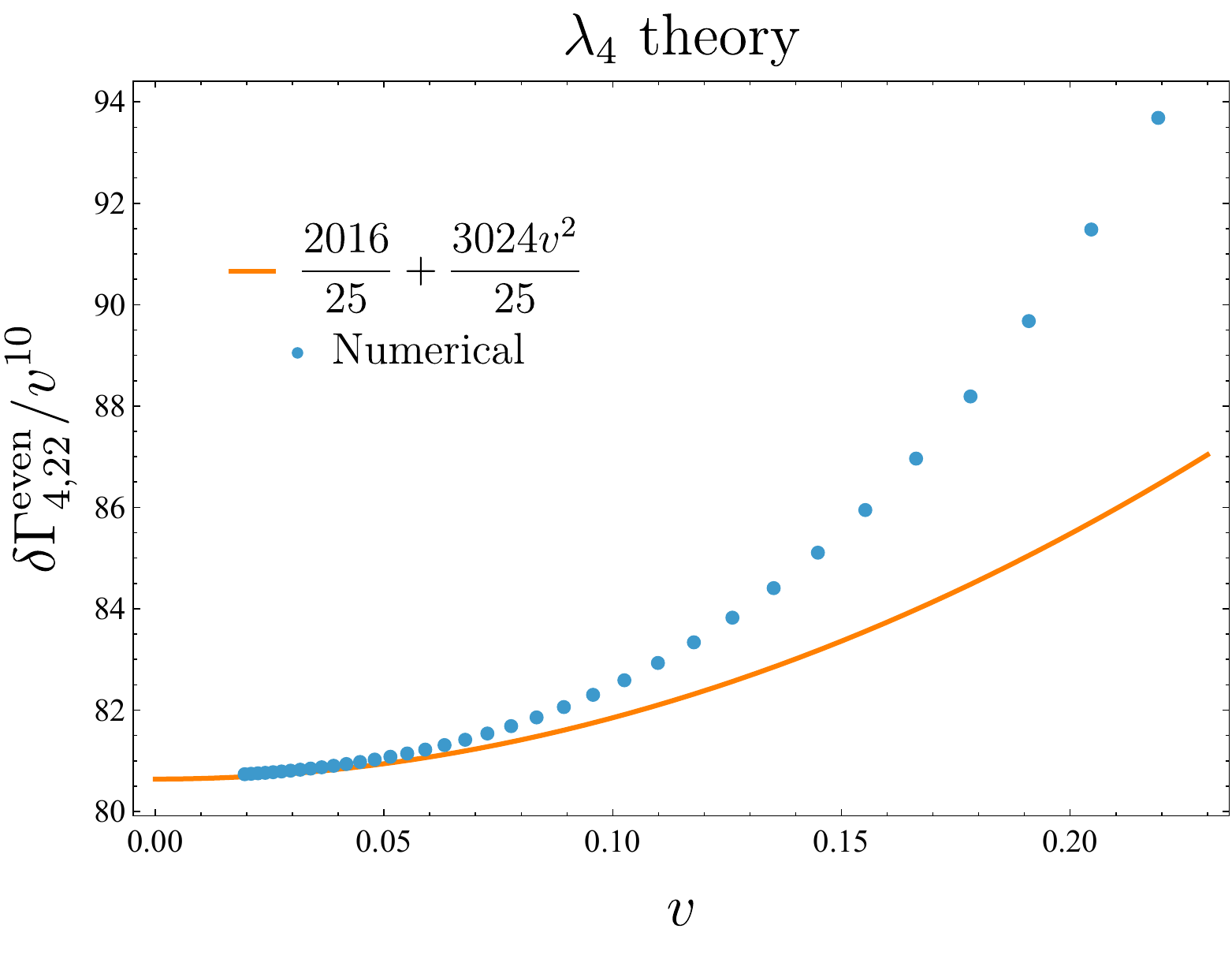}
	\includegraphics[width=0.48\textwidth]{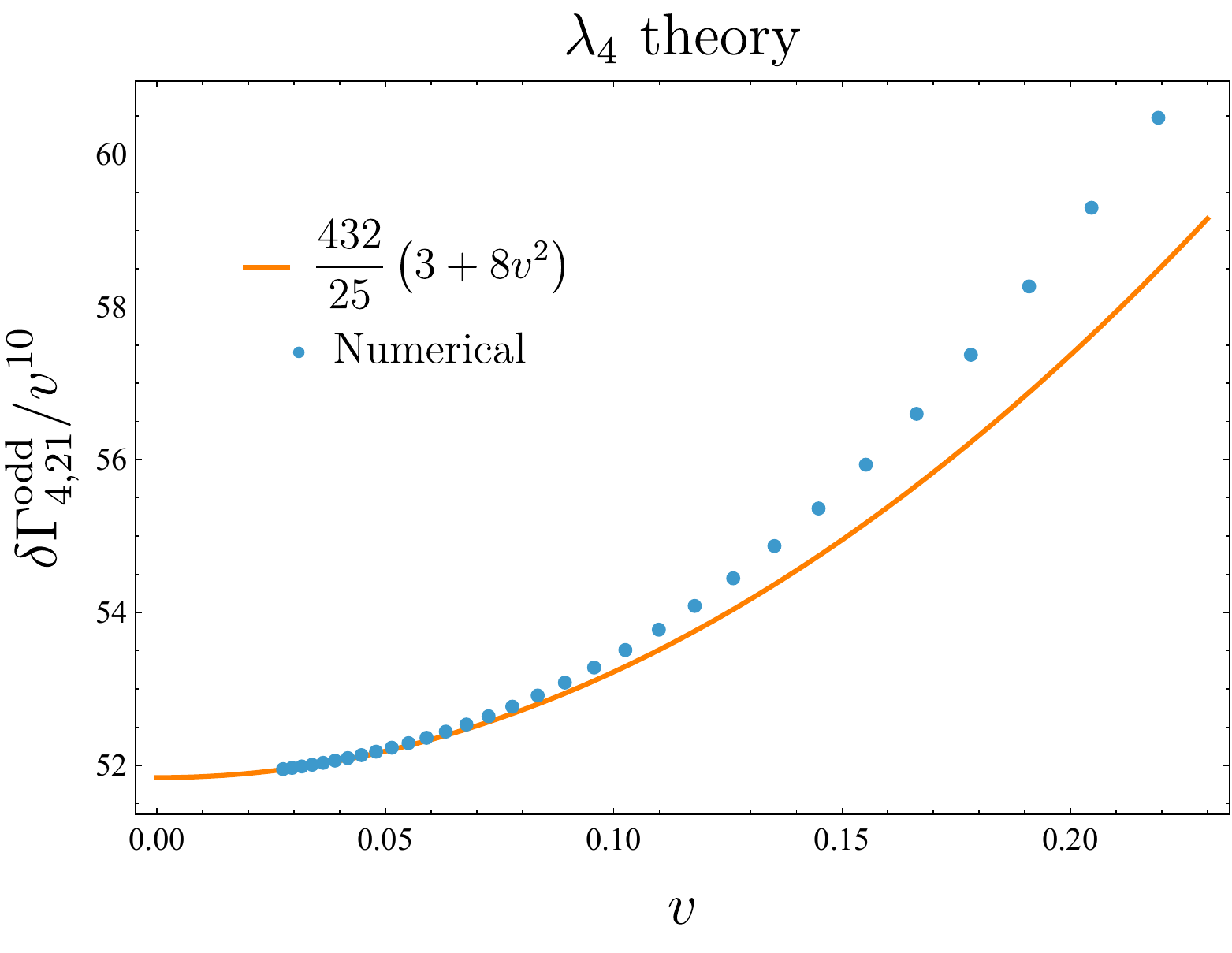}
    \includegraphics[width=0.48\textwidth]{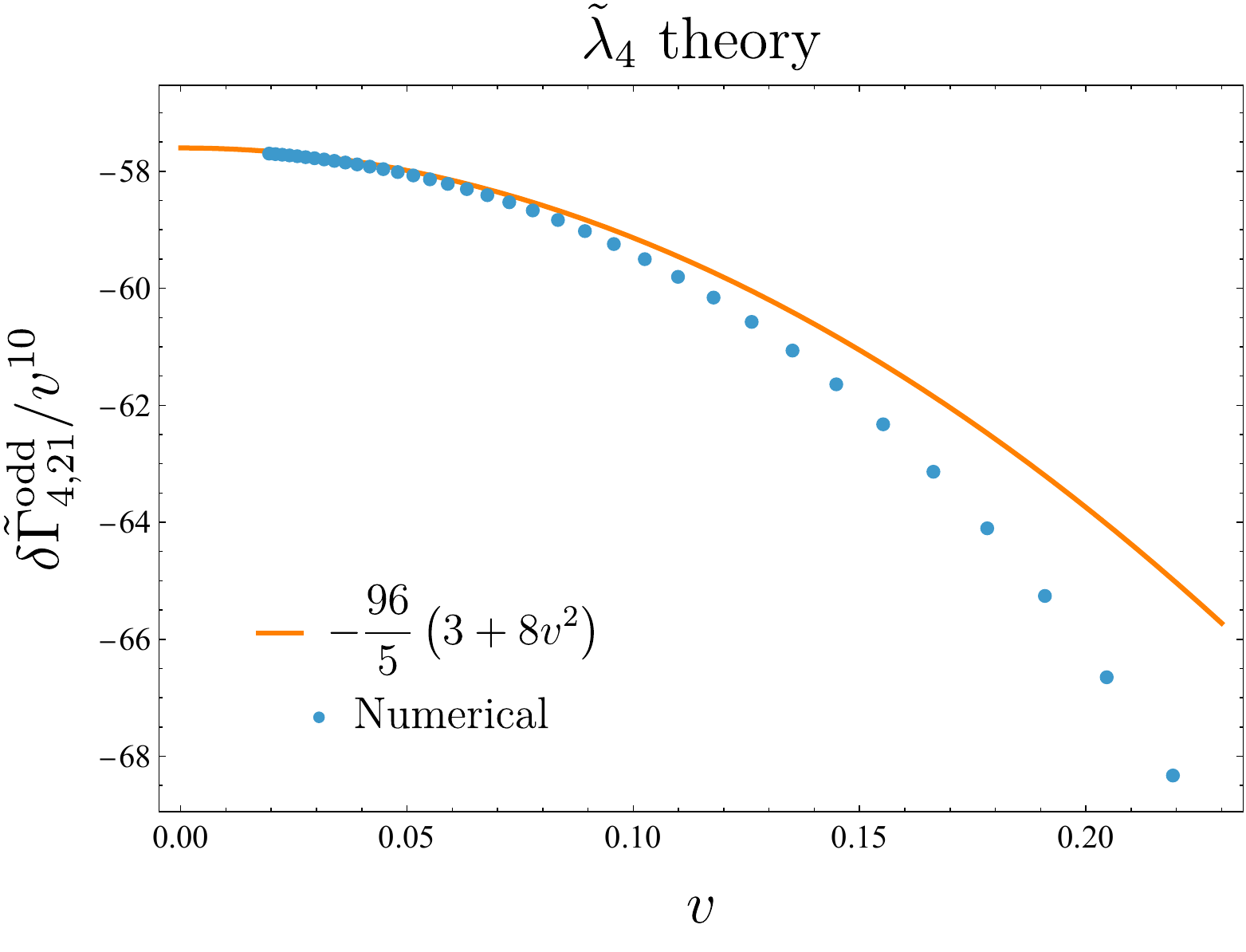}
	\caption{Corrections to the amplitudes $Z_{22}^{\rm even}$ and $Z_{21}^{\rm odd}$ due to cubic higher-derivative corrections. We divide the result by $v^{10}$, showing that the limit for $v\to 0$ is finite. We also show the analytic estimates at 6PN order, showing full agreement with the numerical results for small $v$.}
	\label{figure:correctionsamplitude}
\end{figure}

From the analytic results in \eqref{eq:deltaGamma}, we expect that these coefficients scale as $v^{10}$ for $\ell=2$.\footnote{Except in the case of the even sector for $\tilde\lambda_{4}$ corrections, since $\delta \tilde{\Gamma}^{\rm even}_{4,\ell m}\equiv 0$.} To check this behavior, we plot the ratios $\delta\Gamma^{\rm even}_{2 2}/v^{10}$ and $\delta\Gamma^{\rm odd}_{2 1}/v^{10}$, for each of the theories, in Fig.~\ref{figure:correctionsamplitude}. On top of the numerical data points\footnote{Observe that the numerical results do not go all the way to $v=0$. This is because, for very small $v$, the corrections become extraordinarily tiny and hence they become very challenging to compute, as a huge numerical precision is needed.}, we also show the analytic result in \eqref{eq:deltaGamma} to 6PN order. Clearly, the plots in Fig.~\ref{figure:correctionsamplitude} show that the analytic formulae give the right result for small enough $v$. In particular, this confirms that the leading correction to the amplitude enters at 5PN and that it is due to the tidal Love numbers of the black hole.

For $\ell>2$ (we checked $\ell=3$ and $\ell=4$), our numerical results confirm that the amplitudes are corrected at a higher PN order.
For cubic gravity, the $\ell\ge 3$ amplitudes are corrected at 6PN order, and the leading coefficient we obtain from the numerical approach agrees with the analytic expression in \eqref{eq:deltaGamma} (we do not include the plots here though).
For the quartic theories, we find that the $\ell=3$ amplitudes are corrected at 7PN order, and this effect is due to the $\ell=3$ tidal Love numbers. For $\ell\ge 4$, the corrections appear at 8PN order instead as the effect of the tidal Love numbers enters at a higher order as $\ell$ is increased.

\begin{figure}[t!]
\center
\includegraphics[width=0.48\textwidth]{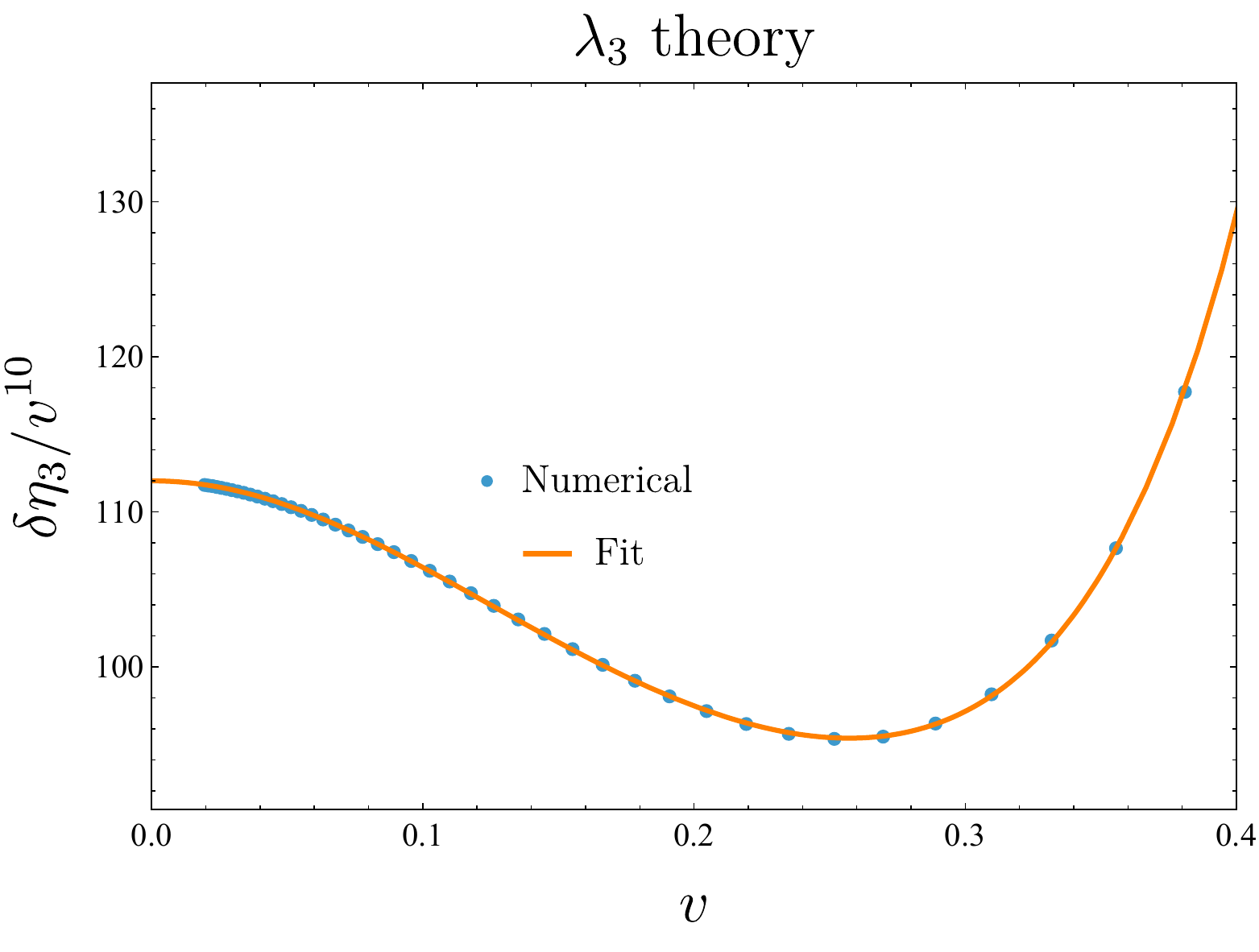}\\
	\includegraphics[width=0.48\textwidth]{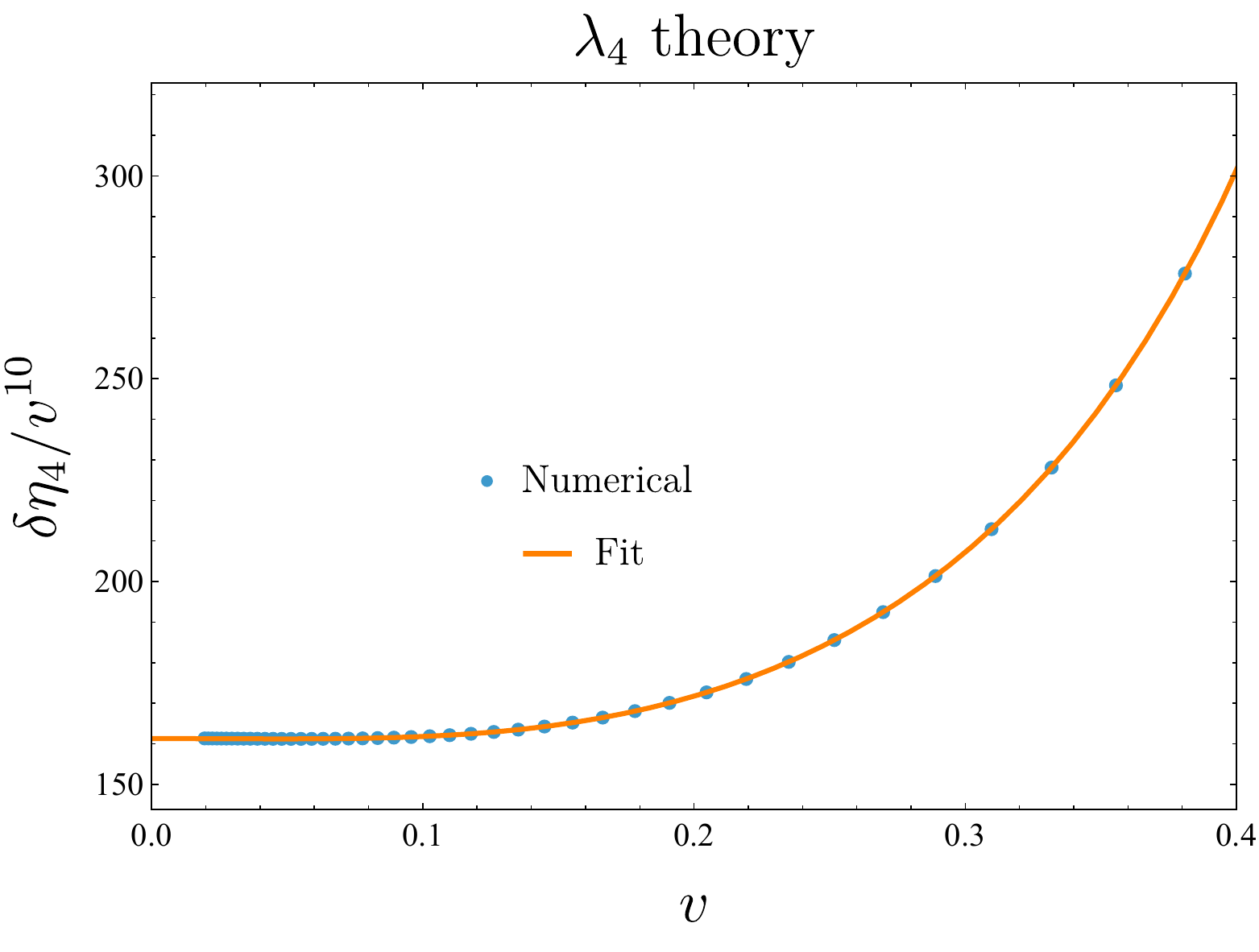}
	\includegraphics[width=0.48\textwidth]{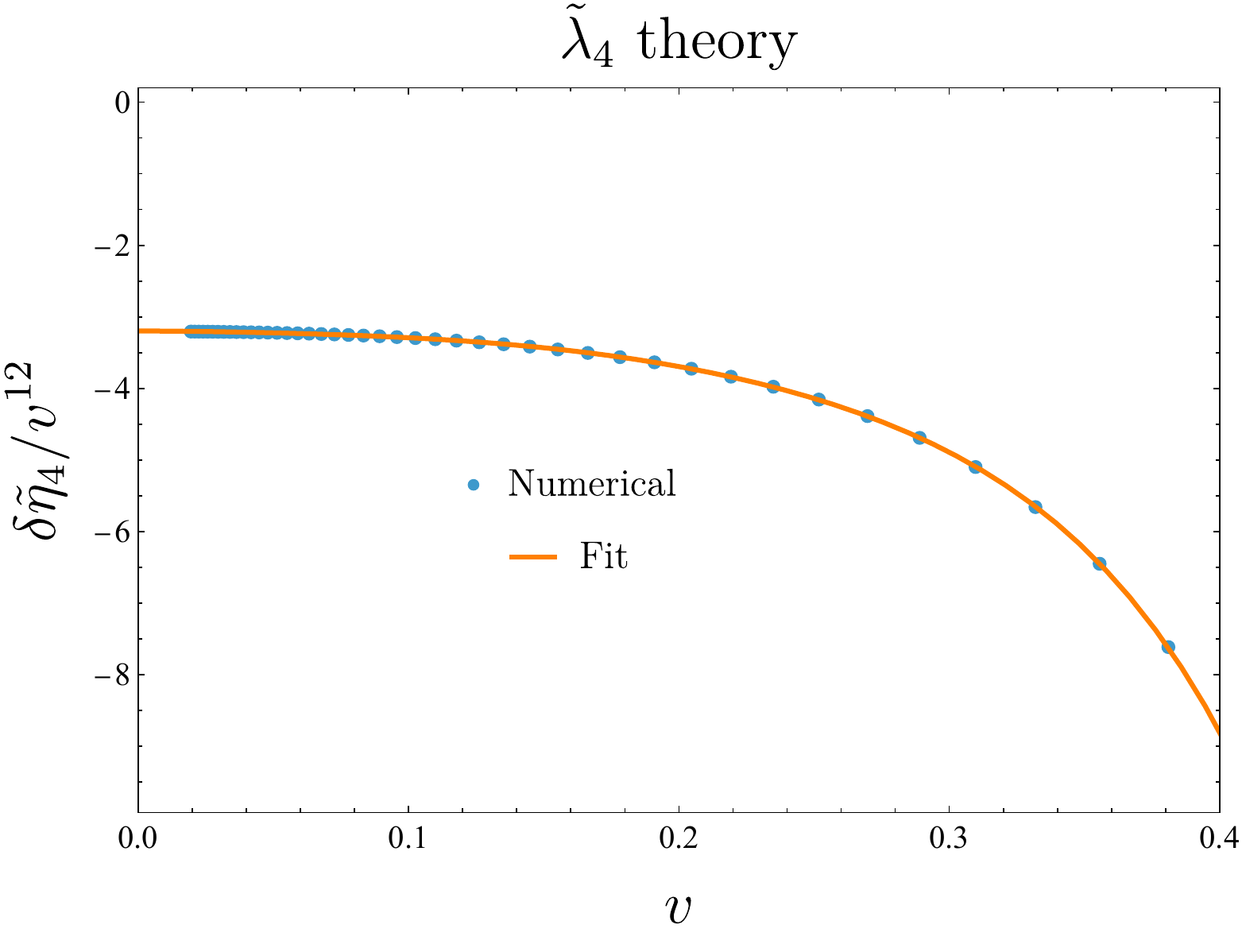}
	\caption{Correction to the energy flux, relative to the GR quadrupole term $\dot{\E}_{N}$, due to cubic corrections and quartic corrections. We divide the result by $v^{10}$ in the $\lambda_{3}$ and $\lambda_{4}$ theories and by $v^{12}$ in the $\tilde{\lambda}_{4}$ case. We show the numerical data points obtained from summing all the modes with $\ell\le 4$ along with a polynomial fit.}
	\label{figure:correctionspower}
\end{figure}

Now we can consider the corrections to the energy flux, given by the coefficient $\delta\eta$ in \eqref{energyflux}. As before, at linear order in the couplings, we expand it as
\begin{equation}
\delta \eta=\frac{\lambda_{3}}{M^4}\delta\eta_{3}+\frac{\lambda_{4}}{M^6}\delta\eta_{4}+\frac{\tilde{\lambda}_{4}}{M^6}\delta\tilde{\eta}_{4}\, ,
\end{equation}
In Fig.~\ref{figure:correctionspower} we show each of these coefficients as a function of $v$. We normalize the result by $v^{10}$ in the $\lambda_{3}$ and $\lambda_{4}$ theories and by $v^{12}$ in the $\tilde{\lambda}_{4}$ case. In fact, in the latter case the corrections to the energy flux start at 6PN instead of 5PN because only the odd sector is modified, as the analytic result in \eqref{deltaeta} shows. In the same figure, we also overlap the numerical data points with a polynomial fit, in which the leading term at $v=0$ is forced to be the analytic value. These fits are given by
\begin{align}
    \delta\eta_{3}^{\rm fit}&=112 v^{10}\left(1-5.853 v^2-2.222 v^3+154.8 v^4-516.2 v^5+624.1 v^6\right)\, ,\\
    \delta\eta_{4}^{\rm fit}&=\frac{4032}{25}v^{10} \left(1-0.4356 v^2+1.999 v^3+76.91 v^4-241.6 v^5+321.2 v^6\right)\, ,\\
    \delta\tilde{\eta}_{4}^{\rm fit}&=-\frac{16}{5} v^{12}
\left(1+4.754 v^2-47.45 v^3+399.1 v^4-1232. v^5+1569.
   v^6\right)\, .
\end{align}
We remark that these fits are not equivalent to a PN expansion (except for the leading term). However, they provide an accurate approximation for the energy flux in the full range of $v$ (up to the ISCO), so they may be used in order to numerically compute inspiral waveforms.

\section{Conclusions}
\label{sec:conclusions}

In this paper we have applied black hole perturbation theory to study the gravitational wave emission in an extreme-mass-ratio inspiral in higher-derivative gravity, in the case of a non-rotating primary black hole. We have derived the master equations governing the linear odd end even metric perturbations in presence of arbitrary sources.
Unlike in GR, these equations cannot be related to equations of confluent Heun type.
 Extending the techniques introduced in \cite{Cipriani:2025ikx}, we found the solutions of the master equations, order by order in  $\omega M$, resumming  the entire $r$ dependence into hypergeometric functions.

The PM solutions describe the GW in the near zone $\omega M\ll \omega r\ll 1$ where the inspiral takes place, but they cannot be extrapolated to the near horizon region where incoming boundary conditions have to be imposed. To circumvent this problem, we solve analytically the master equations in the static limit  and match the incoming solution against its PM expansion in the near zone. The matching determines the Love number describing the tidal response of the geometry to the perturbation, which is non-zero for black holes in higher-derivative gravity. This tidal effect modifies the $\ell$th-multipole of the GW waveform at $(2\ell+1)$ PM order beyond the leading GR result.

Specifying to circular orbits, we computed the corrections (up to 6PN order) to the waveform, the energy and angular momentum fluxes radiated towards infinity.  Our results have been tested against numerical computations.
Our main takeaway is that the leading modification of the GW waveform in the EFT appears at 5PN order and it is proportional to the $\ell=2$ even-parity Love number. In particular, the leading correction to the energy flux is given by \eqref{deltaetaintro}. This formula provides a direct connection between tidal Love numbers and observables, and it confirms that the tidal Love numbers derived in previous literature via different prescriptions \cite{Cardoso:2018ptl,Cai:2019npx,DeLuca:2022tkm,Katagiri:2024fpn,Barbosa:2025uau,Cano:2025zyk,Wang:2026qst} are in fact physically robust quantities. Interestingly, this finite size effect dominates over higher-derivative corrections to the master equations and geodesic motion, which show up at 6PN and 8PN orders for cubic and quartic interactions, respectively. The same conclusion was already noted in \cite{Bernard:2025dyh}.\footnote{We thank Luis Lehner for correspondence on this point.}

It would be interesting to elucidate if previous analyses, like those based on scattering amplitudes \cite{Endlich:2017tqa,Brandhuber:2024bnz},  may have overlooked this tidal effect. Indeed, the fact that finite size effects dominate the EFT corrections to the GW signal may be somewhat counterintuitive, and it is due to an interplay between the two relevant scales in the problem: the size of the black holes and the distance between them. The latter controls the corrections to radiation emission and to geodesic motion, but because higher-derivative corrections decay very fast with distance, these are highly suppressed. On the other hand, the corrections to the individual black hole geometries are much larger, as they are only suppressed with powers of the black hole radius. All in all, this makes the finite size effects become the most important beyond-GR contribution in the EFT extension of GR. For binaries involving rotating black holes, we expect these effects to appear even earlier, at 2PN order, due to the modification of the spin-induced multipole moments \cite{Sennett:2019bpc,Cano:2022wwo}.

Our work contains an implicit assumption given that we are using minimally coupled matter to describe the point particle. However, in the spirit of EFT, the matter Lagrangian can also contain higher-derivative corrections. In fact, if this point particle is to represent a black hole, its worldline action would contain higher-derivative corrections which would be fully fixed by the gravitational action \cite{Gralla:2008fg}. To the best of our knowledge, how to derive such worldline action for the point limit of a black hole in higher-derivative gravity is still an open question. On the other hand, in the gravitational action in \eqref{eq:EFTofGR} we have implicitly performed field redefinitions to remove Ricci-dependent densities \cite{Endlich:2017tqa,Cano:2019ore}. These redefinitions would also affect the matter Lagrangian so that, even if this Lagrangian was minimally coupled in some frame, in general it will not be minimally coupled in the frame of \eqref{eq:EFTofGR} \cite{deRham:2019ctd}. So, for the sake of generality, one should consider an EFT extension of the point particle action as well.
We expect that a modification of the matter source due to metric redefinitions would only introduce contributions starting at 6PN, associated to terms quadratic in the background curvature in the worldline action. Therefore, the tidal interactions, which enter at 5PN, would universally remain the leading beyond-GR effect.

We believe there is much to be understood along these lines. Within the extreme-mass-ratio approximation, it would be interesting to extend our work by analyzing non-circular orbits, considering a more general matter source as we just discussed, and, especially, studying the case of a rotating primary black hole. More broadly, correctly describing all the beyond-GR effects in the case of comparable-mass binaries remains a grand goal.

The computations involved in this research led to very large equations which were impossible to report in paper. For this reason this work comes with an ancillary Mathematica file where the reader can find our complete results.

\section*{Acknowledgments}

We would like to thank Massimo Bianchi, Andreas Brandhuber, Graham Brown, Vitor Cardoso, Giorgio Di Russo, Luis Lehner, David Pere\~niguez, Gabriele Travaglini and Pablo Vives Matasan for useful discussions and comments. The work of P.A.C. is supported by a Ram\'on y Cajal fellowship (RYC2023-044375-I) and by a Proyecto de generaci\'on de conocimiento (PID2024-155685NB-C22) from Spain’s Ministry of Science, Innovation and Universities.
The work of AR is supported by the University of Padua and the Fondazione Cariparo under the STARS@UNIPD 2025 programme (GRASBH -- The gravitational path integral, supersymmetric black holes and higher-derivative corrections).



\appendix
\section{Spherical harmonics}
\label{app:spherical_harmonics}

We follow \cite{Martel:2005ir}. Let $D_{A}$ denote the covariant derivative of the unit-sphere metric $\Omega_{AB}$. The usual scalar spherical harmonics $Y^{\ell m}$ are eigenfunctions of the Laplacian,
 \begin{equation}
 D_{A}D^{A}Y^{\ell m}=-\ell(\ell+1)Y^{\ell m}\, .
 \end{equation}
 They are explicitly given by
 \begin{equation}
 Y^{\ell m}(\theta,\phi)=\sqrt{\frac{(2\ell+1)(\ell-m)!}{4\pi (\ell+m)!}}P_{\ell}^{m}(\cos\theta) e^{im\phi}\, ,
 \end{equation}
 where $P_{\ell}^{m}(x)$ are the associated Legendre functions, and the normalization factor ensures that they satisfy
 \begin{equation}
\int_{0}^{2\pi}d\phi \int_{0}^{\pi} d\theta \sin\theta\, \bar{Y}^{\ell m} Y^{\ell'm'}\equiv \int d\Omega\, \bar{Y}^{\ell m} Y^{\ell'm'}=\delta_{\ell \ell'}\delta_{m m'}\, .
 \end{equation}
 From these we can build vector and tensor spherical harmonics, applying the covariant derivative.
 For the vector harmonics we have an even-parity type $Y_{A}^{\ell m}$ and an odd parity type $X_{A}^{\ell m}$, given by
 \begin{align}
 \label{eq:evenvectorharmonics}
Y_{A}^{\ell m}&=D_{A}Y^{\ell m}\, ,\\
\label{eq:oddvectorharmonics}
X_{A}^{\ell m}&=-\epsilon_{A}^{\,\,\,\, B}D_{B}Y^{\ell m}\, ,
\end{align}
where $\epsilon_{AB}$ is the Levi-Civita tensor of the 2-sphere, such that $\epsilon_{\theta\phi}=+\sin\theta$.
These satisfy the orthogonality relations
\begin{align}
\int d\Omega\, \bar{Y}^{A\, \ell m} Y_{A}^{\ell'm'}&=\ell (\ell+1)\delta_{\ell \ell'}\delta_{m m'}\, ,\\
\int d\Omega\, \bar{X}^{A\, \ell m} X_{A}^{\ell'm'}&=\ell (\ell+1)\delta_{\ell \ell'}\delta_{m m'}\, ,\\
\int d\Omega\, \bar{Y}^{A\, \ell m} X_{A}^{\ell'm'}&=0\, .
\end{align}
There are two traceless tensor harmonics, again one of each parity type,
\begin{align}
\label{eq:eventensorharmonics}
Y_{AB}^{\ell m}&=\left[D_{A}D_{B}+\frac{1}{2}\ell(\ell+1)\Omega_{AB}\right]Y^{\ell m}\, ,\\
\label{eq:oddtensorharmonics}
X_{AB}^{\ell m}&=-\epsilon_{(A}^{\,\,\,\, C}D_{B)}D_{C}Y^{\ell m}\, .
\end{align}
These satisfy
\begin{align}
\int d\Omega\, \bar{Y}^{AB\, \ell m} Y_{AB}^{\ell'm'}&=\frac{1}{2}(\ell-1)\ell (\ell+1)(\ell+2)\delta_{\ell \ell'}\delta_{m m'}\, ,\\
\int d\Omega\, \bar{X}^{AB\, \ell m} X_{AB}^{\ell'm'}&=\frac{1}{2}(\ell-1)\ell (\ell+1)(\ell+2)\delta_{\ell \ell'}\delta_{m m'}\, ,\\
\int d\Omega\, \bar{Y}^{AB\, \ell m} X_{AB}^{\ell'm'}&=0\, .
\end{align}
In the even-parity sector we have the additional structure $\Omega_{AB}Y^{\ell m}$ which is obviously also a tensor harmonic, and it is pointwise orthogonal to $Y_{AB}^{\ell m}$ and $X_{AB}^{\ell m}$.

\newpage
\section{Coefficients $A_0, A_1$}
\label{app:coefficients}

{\small{
\beaq
 && A_{0,{\rm odd}}= \frac{4 J (2 M-r)}{r^3}+\frac{16 J M \left(380 M^2-195 M r-36 r^4 \omega ^2\right)
   \lambda _3}{r^9}\\
 && A_{1,{\rm odd}}=\frac{4 J (2 M-r)}{r^2}+\frac{16 J M^2 (92 M-51 r) \lambda _3}{r^8}\nn\\
&& A_{0,{\rm even}}= Y(\ft{\pi}{2},0)\left[ \frac{2 J^2 (2 M-r) \left(2-2 m^2+\nu \right)}{E r^4 \nu  (2+\nu )}+\frac{2 J m (2 M-r)
   \left(24 M^2+6 M r \nu +r^2 \nu  (2+\nu )\right)}{r^4 (2+\nu ) (6 M+r \nu )^2 \omega
   }\right] \nn\\
   &&+\lambda_3 Y(\ft{\pi}{2},0)\left\{-\frac{48 E M}{(2 M-r) r^6 (6 M+r \nu )^2} \left[90 M^3+2 M r^2 (-9+\nu ) \nu -r^3 (-1+\nu ) \nu +3 M^2 r
   (-14+11 \nu )\right] \right. \nn\\
   && +\frac{8 J^2 M}{E r^{10} \nu  (2+\nu ) (6 M+r \nu )^2} \left[-1368 M^4
   \left(-2+2 m^2-\nu \right)+48 M^3 r \left(-36+m^2 (36-25 \nu )+4 \nu +11 \nu
   ^2\right) \right. \nn\\
  &&   +2 M^2 r^2 \nu  \left(-324+m^2 (360-92 \nu )-88 \nu +37 \nu ^2\right)+3 r^4
   \nu ^2 \left(m^2 (-2+\nu )+2 (2+\nu )\right)   \nn\\
&&   -6 M r^3 \nu ^2 \left(m^2 (-19+\nu )+9 (2+\nu )\right]
   +\frac{8 J m M}{r^{10} (2+\nu ) (6 M+r \nu )^3  \omega }
   \left[39456 M^5+48 M^4 r (-984+563 \nu ) \right. \nn\\
   && \left. +2 M^2 r^3 \nu  \left(6372-3373 \nu +19 \nu ^2-234 r^2 \omega ^2\right)-12 M^3 r^2 \left(-1116+2973 \nu -383 \nu ^2+72 r^2 \omega   ^2\right) \right. \nn\\
   && \left. {+}3 r^5 \nu ^2 \left(2 \nu ^2{-}76{+}2 r^2 \omega ^2{+}\nu  \left(26{-}r^2 \omega   ^2\right)\right)
   {-}12 M r^4 \nu  \left(78{+}15 \nu ^2{+}\nu ^3{-}3 r^2 \omega ^2{+}\nu
  \left(6 r^2 \omega ^2 {-}220\right)\right] \right\} \nn\\
  && A_{1,{\rm even}}= \frac{4 J m (r{-}2 M)^2  Y(\ft{\pi}{2},0) }{r^3 (2{+}\nu ) (6 M{+}r \nu ) \omega } {+}\frac{ \lambda_3 Y(\ft{\pi}{2},0)} {r^5 (6 M+r \nu )} \left[48 E M (3
   M{-}r){+}\frac{48 J^2 M (r{-}2 M)^2 \left(m^2{-}2{-}\nu \right)}{E r^3
   (2{+}\nu )  } \right. \nn\\
  &&\left.\left.   \right. \right. \nn\\
   && \left. \left. {-}\frac{16 J m M (2 M{-}r)}{r^4 (2{+}\nu ) (6 M{+}r \nu ) \omega } \left(84 M^3{+}2 M^2 r (76 \nu{-}27 ){-}3
   r^3 \nu  \left(2 \nu{-}8 {+}r^2 \omega ^2\right){-}6 M r^2 \left(19 \nu{-}6 {-}2 \nu ^2{+}3 r^2
   \omega ^2\right)\right)\right)  \right]\nn
   \eeaq
}}
 \section{$\eta^{\rm GR}$ up to $v^{12}$}\label{app:etaGR}
 \label{etagr}
 For convenience of the reader we display here the 6PN result for the energy flux in GR:
\small{ \beaq
  &&\eta^{\rm GR}  =    1{-}\frac{1247 v^2}{336}{+}4
   \pi  v^3{-}\frac{44711 v^4}{9072}{-}\frac{8191 \pi  v^5}{672}+v^6\left(\frac{6643739519 }{69854400}-\frac{1712\gamma}{105}+\frac{16\pi^2}{3}-\frac{3424\log 2}{105}\right.\nn\\
   &&\left. -\frac{1712\log v}{105}\right){-}\frac{16285 \pi
   v^7}{504}+v^8\left(-\frac{323105549467}{317837520}+\frac{232597\gamma}{4410}-\frac{1369\pi^2}{126}
   +\frac{39931\log 2}{294}-\frac{47385\log 3}{1568}\right.\nn\\
   &&\left.+\frac{232597\log v}{4410}\right)+v^9\left(\frac{265978667519\pi}{745113600}-\frac{6848 \gamma \pi
   }{105}-\frac{13696 \pi\log (2)}{105}-\frac{6848}{105} \pi  \log (v)\right)\nn\\
   &&+v^{10}\left(\frac{916628467 \log (v)}{7858620}-\frac{424223 \pi
   ^2}{6804}+\frac{916628467 \gamma
   }{7858620}-\frac{2500861660823683}{2831932303200}+\frac{47385 \log
   (3)}{196}\right.\nn\\
   &&\left.-\frac{83217611 \log (2)}{1122660}\right)+v^{11}\left(\frac{177293 \pi  \log (v)}{1176}+
  \frac{8399309750401\pi}{101708006400}+\frac{177293 \pi\,\gamma
   }{1176}\right.\nn\\
   &&\left.+\frac{8521283 \pi\log (2)}{17640}-\frac{142155 \pi\log (3)}{784}\right) +v^{12}\left[\frac{1465472 \log
   ^2(v)}{11025}+\left(-\frac{246137536815857}{157329572400}\right.\right.\nn\\
   &&\left.\left.+\frac{2930944
   \gamma }{11025}-\frac{27392 \pi ^2}{315}+\frac{5861888
   \log(2)}{11025}\right) \log (v)-\frac{27392 \zeta (3)}{105}-\frac{256 \pi
   ^4}{45}+\frac{3803225263 \pi ^2}{10478160}\right.\nn\\
   &&\left.+\frac{1465472 \gamma
   ^2}{11025}+\frac{2067586193789233570693}{602387400044430000}+\frac{586188
   8 \log ^2(2)}{11025}-\frac{37744140625 \log
   (5)}{260941824}\right.\nn\\
   &&\left.-\frac{437114506833 \log (3)}{789268480}+\gamma
   \left(-\frac{246137536815857}{157329572400}-\frac{27392 \pi
   ^2}{315}+\frac{5861888 \log (2)}{11025}\right)\right.\nn\\
   &&\left.-\frac{54784}{315} \pi ^2
   \log (2)-\frac{271272899815409 \log (2)}{157329572400}\right]\,.
  \eeaq
  }

\bibliography{refs.bib}
\bibliographystyle{JHEP}

\end{document}